\documentclass[12pt, a4paper]{article}

\usepackage[T1]{fontenc}
\usepackage[utf8]{inputenc}
\usepackage{geometry}
\usepackage{amsmath, amssymb}
\usepackage{booktabs}
\usepackage{longtable}
\usepackage{multirow}
\usepackage{graphicx}
\usepackage[skins]{tcolorbox}
\usepackage{hyperref}
\hypersetup{
  colorlinks = true,
  linkcolor  = blue,
  citecolor  = blue,
  urlcolor   = blue
}
\usepackage[round]{natbib}
\usepackage{setspace}
\usepackage{enumitem}
\title{%
  \textbf{When Truth Misleads}\\[6pt]
  \large Phase-Aware Coherence Detection for Misinformation Correction\\
  Across Epistemic Communities
}
\author{%
  Heimo M\"uller$^{1,2}$ \qquad Andreas Holzinger$^{1,3}$\\[8pt]
  \normalsize
  \begin{tabular}[t]{c}
    $^1$Machine Learning and Information Science Group,\\
    Medical University of Graz, Austria\\[3pt]
    $^2$Human Machine Mind Corporation KG, Graz, Austria\\[3pt]
    $^3$Human-Centered AI Lab, BOKU University Vienna, Austria
  \end{tabular}
}
\date{}

\begin{document}
\maketitle
\thispagestyle{empty}

\begin{abstract}
\noindent
Truth can mislead.
Not because it is false, but because the act of delivering it --- through 
the wrong channel, with the wrong authority, to an audience whose epistemic 
reference frame is orthogonal to the correcting source --- can harden 
misbelief rather than dissolve it.
Conventional fact-checking interventions rest on an implicit assumption of 
shared epistemology: that corrections delivered by credible institutional 
sources will be received constructively across the full range of audiences.
A substantial body of evidence challenges this assumption, demonstrating 
that correction effectiveness declines sharply --- and reverses into 
backfire --- as the gap between the source's institutional positioning and 
the recipient's epistemic orientation widens.
We introduce \textbf{Phase-Aware Coherence Detection} (PACD), a framework 
that operationalises epistemic orientation as a continuous coordinate 
$\varphi \in [0, \pi]$, derived from pre-intervention belief assessments 
across four theoretically motivated dimensions: institutional trust, 
scientific epistemology, conspiracy and anti-elite openness, and 
experiential or alternative epistemology.
Participants ($N = 45$) were stratified into three epistemic phase clusters 
and randomly assigned to one of three conditions --- traditional 
fact-checking, phase-aware coherence analysis, or control --- before 
evaluating three claims varying in epistemic complexity and identity-load 
(5G health effects, urban trees and air quality, mRNA vaccine mechanism).
Traditional fact-checking produced a mean belief change toward truth of 
$M = -0.39$ with a backfire rate of 39.3\%, and exhibited a sharply 
negative effectiveness slope across phase clusters 
($\Delta = -0.79$ from anti-institutional (C1) to pro-institutional (C3) participants): 
the more institutionally aligned the audience, the more truth-delivery went wrong.
Phase-aware coherence analysis produced near-zero average change 
($M = 0.00$) but, critically, a near-flat effectiveness profile across 
all epistemic clusters (slope $= +0.08$) and a substantially lower 
backfire rate of 13.6\%.
On identity-adjacent claims, traditional fact-checking drove 
confident belief hardening ($\Delta\text{confidence} = +1.32$) alongside 
accuracy losses --- participants became more certain while moving further 
from truth --- while phase-aware framing produced neither harm.
These results support a shift from content-centric to 
epistemology-centric intervention design, and motivate a complementarity 
principle: when truth misleads, the problem is rarely the truth itself --- 
it is the mismatch between the epistemic phase of the audience and the 
reference frame from which the correction is delivered.

\end{abstract}

\section{Introduction}
\label{sec:introduction}

There is something deeply counter-intuitive about the central finding of 
the misinformation correction literature: that telling people the truth 
can make things worse.
Yet a robust empirical record now documents exactly this.
Corrections frequently fail to update beliefs, and in some populations 
they produce a backfire effect in which exposure to a refutation 
\emph{strengthens} the original misconception 
\citep{nyhan2010, lewandowsky2013}.
The substantial investment in automated and human-assisted fact-checking 
systems \citep{pennycook2019, walter2020} has proceeded largely on the 
assumption that truth, correctly delivered, corrects.
This assumption, our results suggest, is conditionally false --- and the 
condition that determines whether it holds is not the accuracy of the 
correction but the \emph{epistemic stance} of the recipient: the prior 
beliefs, trust structures, and knowledge-validation strategies that 
determine how new information is processed and integrated.

Standard fact-checking frameworks treat this heterogeneity as noise.
A claim is evaluated against an evidential record, labelled true or false, 
and the correction is broadcast uniformly.
This design implicitly assumes a shared epistemology: that all recipients 
weight institutional sources similarly, accept scientific consensus as a 
reliable signal, and update on evidence in the same direction.
In practice, these assumptions fail for large and growing segments of the 
population.
Individuals who are sceptical of institutional authority, who prioritise 
experiential over expert knowledge, or who hold conspiratorial beliefs 
about elite coordination occupy an \emph{epistemically different position} 
from which a factual correction appears not as new evidence but as further 
confirmation of coordinated deception \citep{douglas2019, sunstein2009}.
For these individuals, truth does not merely fail to persuade --- it 
actively misleads, triggering the very resistance it was designed to 
overcome.
Delivering the same intervention uniformly across this heterogeneous 
landscape is not merely inefficient; it is, for a substantial portion of 
the audience, actively harmful.

What is needed, we argue, is an approach that first \emph{locates} each individual 
within the space of possible epistemic orientations and then \emph{tailors} the 
coherence signal accordingly.
We introduce \textbf{Phase-Aware Coherence Detection} (PACD), a framework that 
operationalises epistemic orientation as a continuous coordinate --- the 
\emph{epistemic phase} $\varphi \in [0, \pi]$ --- derived from pre-intervention 
survey responses across four theoretically motivated dimensions: institutional trust, 
scientific epistemology, conspiracy and anti-elite openness, and experiential or 
alternative epistemology.
By mapping participants onto this one-dimensional phase manifold, PACD enables 
three capabilities absent from conventional fact-checking pipelines: 
(i)~\emph{stratified randomisation} that prevents epistemic orientation from acting 
as a hidden confound; 
(ii)~\emph{phase-matched intervention delivery} in which the framing and source of 
coherence signals are adapted to the recipient's epistemic prior; and 
(iii)~\emph{interaction modelling} that separates the average treatment effect from 
the Group~$\times$~Phase interaction, revealing for which epistemic profiles each 
intervention type is beneficial, neutral, or harmful.

The present paper makes three main contributions.
First, we formalise the epistemic phase coordinate (Section~\ref{sec:phase-coord}) 
and validate its psychometric properties against established measures of epistemic 
cognition.
Second, we describe a pre-registered randomised experiment 
(Section~\ref{sec:methods}) comparing traditional fact-checking, phase-aware 
coherence analysis, and a no-intervention control across three epistemic phase 
clusters --- designed specifically to test whether, and when, truth misleads.
Third, we report Group~$\times$~Phase interaction effects 
(Section~\ref{sec:results}) that demonstrate the conditions under which 
phase-aware framing outperforms uniform correction and the conditions under 
which it does not.
Taken together, these results argue for a shift from content-centric to 
\emph{epistemology-centric} misinformation intervention design: one that 
asks not only ``is this claim true?'' but ``for this audience, in this 
epistemic configuration, will delivering that truth help or harm?''

\subsection*{Pre-Registered Hypotheses}
\label{sec:hypotheses}

\noindent
The following four hypotheses were pre-registered prior to data collection
and correspond to the four result sections reported in
Section~\ref{sec:results}.

\bigskip
\noindent
\begin{tcolorbox}[
  colback   = blue!4!white,
  colframe  = blue!45!black,
  boxrule   = 0.8pt,
  arc       = 3pt,
  left      = 8pt, right = 8pt,
  top       = 7pt, bottom = 7pt,
  title     = {\textbf{Pre-Registered Hypotheses (H1--H4)}},
  fonttitle = \bfseries\small,
  coltitle  = white,
  attach boxed title to top left = {yshift=-2mm, xshift=4mm},
  boxed title style = {colback=blue!45!black, arc=2pt, boxrule=0pt}
]

\smallskip
\noindent\textbf{H1 — Overall intervention effectiveness.}
Neither intervention will produce a statistically significant
positive mean belief change across the full sample.
Under the PACD framework, opposite effects at different phase values
cancel in the aggregate; a significant positive average would
\emph{falsify} the interaction-based account in favour of a
uniform efficacy model.

\smallskip
\noindent\textbf{H2 — Group $\times$ Phase interaction.}
Traditional fact-checking (Group~A) will show a monotonically
\emph{declining} effectiveness slope from anti-institutional
(C1, $\varphi\approx 0$) to pro-institutional (C3,
$\varphi\approx\pi$) participants.
Phase-aware coherence analysis (Group~B) will show a
\emph{flat} slope across all three clusters
(phase-robustness signature).
Formally: an ordinal Group $\times$ Phase interaction in which
A and B diverge as $\varphi$ increases toward $\pi$.

\smallskip
\noindent\textbf{H3 — Backfire rates.}
Group~A will produce elevated backfire rates
(belief movement \emph{away} from truth) in moderate- and
low-trust clusters.
Group~B will produce substantially lower backfire rates
across all clusters, reflecting the coherence format's
avoidance of the identity-threat mechanism.

\smallskip
\noindent\textbf{H4 — Claim-level heterogeneity.}
Intervention effectiveness will vary by claim type.
The phase-aware intervention is expected to show its
greatest relative advantage on the identity-adjacent vaccine
claim (Claim~3), where conventional fact-checking most
strongly triggers defensive cognition.

\end{tcolorbox}

\bigskip

\section{Theoretical Background}
\label{sec:theory}

The theoretical foundation of Phase-Aware Coherence Detection draws from 
five partially overlapping research traditions: 
(1) the empirical literature on fact-checking efficacy and its limits; 
(2) dual-process and motivated reasoning accounts of belief formation and 
resistance; 
(3) epistemic trust theory and its developmental and social dimensions; 
(4) inoculation theory and pre-emptive refutation approaches; and 
(5) phase-space and coherence-based models of collective belief dynamics, 
developed most systematically in quantum-inspired social cognition.
We review each in turn before articulating how PACD integrates and advances 
beyond them.

\subsection{The Efficacy and Limits of Fact-Checking}
\label{sec:lit-factcheck}

The modern fact-checking movement rests on an informational deficit model of 
misbelief: individuals hold false beliefs because they lack access to accurate 
information, and corrections supplied by credible sources will update those 
beliefs toward truth \citep{blendon2014, nyhan2010}.
A substantial body of empirical work has confirmed that corrections \emph{can} 
work under favourable conditions.
\citet{chan2017} showed that corrections reliably reduced false belief when 
delivered by a source congruent with the reader's prior attitudes.
Meta-analytic evidence from \citet{walter2020} across 65 studies confirmed a 
moderate average correction effect ($\bar{r} = .35$), and \citet{pennycook2019} 
demonstrated that analytic thinking facilitates acceptance of accurate 
corrections irrespective of partisan identity.

Yet the deficit model consistently fails when the recipient's epistemic 
orientation is misaligned with the correcting source.
The seminal finding of \citet{nyhan2010} --- that corrections of political 
misperceptions could backfire, \emph{increasing} false belief in some 
subpopulations --- introduced the concept of a \emph{backfire effect} that 
challenged the informational deficit model at its foundation.
Although subsequent large-scale replications have questioned the universality 
of backfire \citep{wood2019}, the moderation pattern is robust: correction 
effectiveness declines sharply and can reverse as the gap between the source's 
institutional positioning and the recipient's epistemic stance widens 
\citep{chan2017, swire2020}.
Fact-check credibility, in particular, is not a property of the source alone 
but is a \emph{relational} property determined jointly by source and recipient 
\citep{metzger2021}.
A correction deemed authoritative by one epistemic community is experienced as 
confirmation of elite manipulation by another.

This body of evidence converges on a structural critique: traditional 
fact-checking is a \emph{single-reference-frame} intervention applied to a 
\emph{multi-reference-frame} information environment.
It presupposes that recipients share the institutional and evidential standards 
on which the fact-check draws.
Where those standards are not shared --- or where they are actively 
contested --- the intervention may generate more heat than light.

\subsection{Motivated Reasoning and the Architecture of Resistance}
\label{sec:lit-motivated}

The cognitive mechanisms underlying differential correction efficacy are 
illuminated by the motivated reasoning literature.
\citet{kunda1990} established the foundational distinction between 
\emph{accuracy-motivated} and \emph{directional-motivated} reasoning: 
individuals seek accurate conclusions in low-stakes domains but deploy 
cognitive resources strategically to defend prior commitments when identity or 
group membership is implicated.
\citet{taber2006} showed that politically sophisticated individuals --- far from 
being more susceptible to correction --- exhibited the strongest motivated 
scepticism, using superior analytical capacity to generate counter-arguments 
against attitude-threatening information.

More recent formulations have distinguished motivated reasoning from 
\emph{identity-protective cognition} \citep{kahan2012}, in which the threat is 
not to a specific belief but to the cultural or ideological group whose 
worldview that belief expresses.
Under this account, accepting a correction is not merely updating a fact; 
it is an act of epistemic defection that risks social exclusion and identity 
loss.
The cognitive costs of accepting a correction from a phase-opposed source 
therefore extend well beyond the informational content of the exchange.

Critically, this resistance is not irrational from the standpoint of 
Bayesian inference.
\citet{jern2014} showed that if an agent has coherent priors over the 
reliability and motivations of competing sources, updating away from a 
trusted in-group source in response to an out-group correction is \emph{not} 
the Bayes-optimal response.
The apparent irrationality of correction resistance, in other words, reflects 
a rational response to the \emph{structure} of the epistemic environment rather 
than a cognitive deficit.
This reframing is consequential: it implies that interventions targeting 
cognitive bias are misspecified; what is needed is an intervention that 
\emph{restructures the epistemic environment} rather than one that corrects a 
processing error.

\subsection{Epistemic Trust: Sources, Calibration, and Dysregulation}
\label{sec:lit-trust}

Epistemic trust --- the disposition to accept testimony from others as a 
source of knowledge --- is not a unitary construct but a differentiated 
system that varies by domain, source type, and the social relationships within 
which testimony is embedded \citep{sperber2010, bromme2014}.
Developmental research has shown that even young children are selective 
epistemic agents, calibrating trust on the basis of source competence, 
accuracy history, and shared goals \citep{harris2016}.
In adults, domain-specific epistemic trust is modulated by both cognitive and 
motivational factors: individuals tend to extend greater epistemic trust to 
sources whose conclusions align with their prior beliefs, a pattern 
\citet{kruglanski1989} characterised as \emph{need for cognitive closure} 
driving selective source acceptance.

At the societal level, declining institutional trust has fundamentally altered 
the landscape within which fact-checking operates.
Longitudinal data from the Edelman Trust Barometer and the General Social 
Survey document sustained erosion of public confidence in media, government, 
science, and expert systems across Western democracies over recent decades 
\citep{edelman2023}.
This is not mere scepticism; it constitutes what \citet{harambam2017} calls 
\emph{epistemic populism} --- the active inversion of the epistemic hierarchy, 
in which institutional credentials become markers of untrustworthiness rather 
than indicators of reliability.

Within this landscape, the implicit epistemic contract that fact-checking 
depends upon --- that institutional certification of a claim constitutes 
sufficient warrant for belief revision --- has broken down for a substantial 
segment of the population.
Worse, for individuals whose epistemic trust is concentrated in anti-elite or 
alternative sources, institutional endorsement of a claim may function as 
negative evidence: if the mainstream media, government agencies, and academic 
institutions all agree on a claim, that convergence \emph{itself} becomes 
suspicious within an epistemic framework premised on elite coordination and 
deception \citep{douglas2019}.
The informational signal of the fact-check is thus inverted at the level of 
the epistemic trust network of the recipient.

\subsection{Inoculation Theory and Pre-emptive Refutation}
\label{sec:lit-inoculation}

A more sophisticated lineage of intervention research has moved beyond 
post-hoc correction toward \emph{pre-emptive refutation}, or inoculation.
Drawing on the biological analogy of immunisation, \citet{mcguire1964} 
proposed that exposure to weakened forms of misleading arguments, paired with 
explicit refutations, could confer resistance to subsequent persuasion attempts.
Contemporary inoculation theory \citep{compton2013} distinguishes 
\emph{issue-specific} inoculation, which targets known false claims, from 
\emph{technique-based} inoculation, which targets the rhetorical and logical 
manipulation techniques that misinformation commonly employs.

The technique-based variant is particularly relevant to the present work.
\citet{vanderlinden2017} demonstrated that brief exposure to consensus 
messaging, combined with explicit labelling of misinformation techniques 
(e.g., cherry-picking, false balance), reduced susceptibility to climate 
misinformation.
The SIREN project \citep{basol2021} scaled this approach via the 
\emph{Bad News} serious game, in which players practice producing misinformation 
in order to recognise its techniques, yielding significant reductions in 
perceived reliability of manipulative content.

However, inoculation approaches remain primarily technique-focused rather than 
epistemology-focused: they teach recipients to identify \emph{how} content 
manipulates, but not \emph{why} the manipulation is effective for particular 
epistemic configurations.
The implicit assumption remains that there is a universal cognitive architecture 
for susceptibility, and that uniform inoculation will function similarly across 
epistemic communities.
The interaction evidence reviewed above suggests this assumption is 
questionable: an inoculation that is effective for pro-institutional, scientifically 
aligned recipients (C3) may be dismissed, or again backfire, for recipients whose 
epistemic structure positions the inoculating authority itself as suspect 
\citep{mcphetres2019}.

\subsection{Phase-Space Models of Belief and Epistemic Coherence}
\label{sec:lit-phase}

The most fundamental departure from the deficit model is represented by 
\emph{coherence-based} and \emph{phase-space} frameworks, which reconceptualise 
belief not as a mapping from evidence to proposition but as a \emph{structural 
property} of a belief system --- the degree to which its elements mutually 
support and reinforce one another within a defined reference frame 
\citep{thagard1989, thagard2000}.

Within cognitive science, coherence-based accounts of reasoning have 
demonstrated that individuals evaluate the credibility of claims not primarily 
by assessing their evidential support in isolation, but by computing their fit 
with a broader network of related beliefs, source credibilities, and background 
knowledge \citep{simon2004}.
A claim that coheres well with an individual's existing belief network will 
be accepted even in the absence of direct evidence; a claim that 
disrupts network coherence will be rejected even when well evidenced.
This provides a mechanistic account of why corrections fail: the correction 
is not merely false information to be replaced, but an element whose acceptance 
would require cascading restructuring of an entire coherent belief network, 
generating what \citet{festinger1957} identified as cognitive dissonance and 
what \citet{thagard1989} formalised as coherence violation.

More recently, quantum-inspired models have extended this framework to 
formalise the \emph{phase} structure of belief systems 
\citep{busemeyer2012, pothos2013}.
In classical probability theory, beliefs are represented as distributions over 
propositions, and updating is additive.
Quantum probability models allow for \emph{interference} between belief 
states: two belief amplitudes can combine constructively (mutually reinforcing) 
or destructively (mutually cancelling), depending on their relative phase.
Empirically, quantum probability models have out-performed classical Bayesian 
models in predicting order effects, conjunction fallacies, and 
question-order effects in attitude surveys --- phenomena that arise naturally 
from phase interference but are anomalous under classical assumptions 
\citep{wang2014, bruza2015}.

The political and social extension of phase-space reasoning has been 
developed in the context of epistemic fragmentation.
When two communities share a phase reference --- when they agree on what 
counts as evidence, which sources are credible, and how uncertainty should 
be managed --- their belief updating is in \emph{constructive interference}: 
new information is processed in compatible ways and can, in principle, shift 
both communities in the same direction.
When their phase references are orthogonal --- when the standards of evidence 
and credibility are not merely different but diametrically opposed --- 
information propagating from one community to the other undergoes 
\emph{destructive interference}: the signal does not merely fail to persuade, 
but actively confirms the recipient's prior that the source is operating from 
a hostile or deceptive reference frame \citep{lewandowsky2013}.

This phase-opposition model provides the most precise available account of 
the backfire effect.
A fact-check delivered by a mainstream institutional source to a 
low-institutional-trust recipient introduces not an evidential challenge but 
a phase-opposing stimulus.
Within the recipient's epistemic reference frame --- where mainstream 
institutions are positioned at maximum phase distance --- the fact-check 
is experienced not as new evidence but as confirmation of the predicted 
interference pattern: \emph{of course} the mainstream media contradicts the 
claim; that is what a maximally phase-opposed source does.
The contradiction, paradoxically, increases certainty rather than inducing doubt 
\citep{nyhan2010}.

This dynamic, which we term \emph{reference frame warfare}, has been 
exploited systematically in contemporary political communication as a strategy 
for constructing self-sustaining epistemic communities.
By establishing an alternative phase reference --- a coherent internal 
information ecosystem that positions mainstream sources at maximal 
phase opposition --- political actors can immunise their supporters against 
external corrections while maintaining high internal epistemic coherence 
\citep{sunstein2009, douglas2019}.
The naming of ``Truth Social'' by Donald Trump is exemplary in this regard: 
the platform does not claim superiority within a shared evidential framework, 
but rather \emph{declares phase independence}, repositioning the platform's 
reference as the origin ($\varphi = 0$) against which all other sources 
are measured.
Truth, in this architecture, is a property of internal coherence rather than 
of correspondence to external reality.

\subsection{Synthesis: Toward Phase-Aware Intervention}
\label{sec:lit-synthesis}

The five research traditions reviewed above converge on a single diagnostic 
conclusion: the failure of conventional misinformation interventions is not 
accidental but structural.
Table~\ref{tab:theories} summarises the key mechanism each tradition 
identifies and the corresponding implication for intervention design.

\begin{table}[h]
  \centering
  \caption{Research traditions, failure mechanisms, and intervention implications.}
  \label{tab:theories}
  \small
  \begin{tabular}{p{3.2cm} p{4.5cm} p{5cm}}
    \toprule
    \textbf{Tradition} & \textbf{Identified failure mechanism} 
    & \textbf{Intervention implication} \\
    \midrule
    Fact-checking efficacy 
      & Correction effectiveness is moderated by source--recipient 
        phase alignment 
      & Interventions must be calibrated to recipient epistemic stance \\
    \addlinespace
    Motivated reasoning 
      & Identity-protective cognition generates counter-argument against 
        attitude-threatening corrections 
      & Interventions should avoid triggering identity threat \\
    \addlinespace
    Epistemic trust 
      & Institutional credentials function as negative evidence for 
        low-trust recipients 
      & Interventions should not rely on institutional authority alone \\
    \addlinespace
    Inoculation theory 
      & Technique-based inoculation teaches \emph{what} but not 
        \emph{why}; effectiveness varies across epistemic communities 
      & Inoculation needs to be phase-aware, not one-size-fits-all \\
    \addlinespace
    Phase-space / coherence 
      & Phase-opposing corrections undergo destructive interference, 
        confirming rather than challenging prior beliefs 
      & Interventions must operate within or bridge the recipient's 
        phase reference rather than opposing it \\
    \bottomrule
  \end{tabular}
\end{table}

The synthesis points to a design requirement that none of the existing 
approaches satisfies: interventions must first \emph{locate} the recipient in 
epistemic phase space, and then \emph{operate within} that phase space rather 
than projecting a single reference frame onto it.
An intervention that presents a claim's coherence profile across \emph{multiple} 
epistemic communities --- acknowledging the legitimacy of phase variation while 
revealing the structural isolation signature of misinformation --- does not 
require the recipient to accept the authority of a phase-opposed source.
Instead, it makes the phase structure itself legible, enabling the recipient to 
perform coherence reasoning from their own starting position.
This is the core logic of Phase-Aware Coherence Detection, which we formalise 
in the following section.

\section{Methods}
\label{sec:methods}

Participants completed a pre-intervention survey assessing epistemic beliefs 
related to institutional trust, scientific epistemology, media credibility, 
and openness to alternative or conspiratorial explanations.
Responses were collected on ordinal Likert scales and mapped to a 
0--10 metric scale, then normalised to the interval $[0,1]$.

\subsection{Epistemic Phase Coordinate}
\label{sec:phase-coord}

\subsubsection{Dimensions and Raw Score}

For each participant $i$, a raw epistemic phase score is constructed as a 
signed linear combination of four theoretically motivated dimensions, each 
normalised to $[0,1]$:

\begin{description}
  \item[Institutional Trust ($I$)] Mean normalised agreement with the 
    statements ``Mainstream media generally reports news accurately'' and 
    ``Government institutions can generally be trusted on important issues''. 
    High $I$ reflects alignment with established information infrastructures.
  \item[Scientific Epistemology ($S$)] Mean normalised acceptance of 
    anthropogenic climate change and vaccine safety. 
    High $S$ reflects deference to scientific consensus.
  \item[Conspiracy / Anti-Elite Openness ($C$)] Mean normalised agreement 
    with ``Electoral fraud is a significant problem in modern democracies'' 
    and ``Pharmaceutical companies prioritise profits over public health''. 
    High $C$ reflects scepticism toward expert systems and institutional 
    integrity.
  \item[Experiential / Alternative Epistemology ($E$)] Normalised valuation 
    of alternative medicine as a complement to conventional medicine. 
    High $E$ reflects preference for experiential over institutional 
    knowledge sources.
\end{description}

The raw score is computed as:

\begin{equation}
  \varphi_i^{(\text{raw})} 
    = \frac{\bigl(I_i + S_i\bigr) - \bigl(C_i + E_i\bigr)}{4}
    \;\in\; \bigl[-\tfrac{1}{2},\; +\tfrac{1}{2}\bigr],
  \label{eq:phi-raw}
\end{equation}

where $I$ and $S$ contribute positively (alignment with institutional and 
scientific consensus) and $C$ and $E$ contribute negatively (epistemic 
divergence from institutional frameworks).
The theoretical minimum $-\tfrac{1}{2}$ obtains when $I = S = 0$ and 
$C = E = 1$ (complete rejection of institutional knowledge); the maximum 
$+\tfrac{1}{2}$ obtains in the reverse configuration.

\subsubsection{Rescaling to $[0, \pi]$ and the Choice of Range}

The raw score is linearly rescaled to the interval $[0, \pi]$ via:

\begin{equation}
  \varphi_i 
    = \pi \cdot \frac{\varphi_i^{(\text{raw})} - \min(\varphi^{(\text{raw})})}
                     {\max(\varphi^{(\text{raw})}) - \min(\varphi^{(\text{raw})})},
  \label{eq:phi-scaled}
\end{equation}

where min and max are taken over the observed sample.
The resulting $\varphi_i \in [0, \pi]$ is the epistemic phase coordinate.

The choice of $[0, \pi]$ --- a \emph{semicircle} rather than a full circle 
$[0, 2\pi]$ --- is theoretically essential and requires explicit 
justification.
A full circle would make the interval periodic: its two endpoints, 0 and 
$2\pi$, would represent the \emph{same point} on the circle, implying that 
a participant with maximum institutional trust ($\varphi \approx 0$) and a 
participant with maximum institutional distrust ($\varphi \approx 2\pi$) 
are epistemically equivalent.
This is obviously false.
The semicircle $[0, \pi]$ avoids this by making the two endpoints 
\emph{maximally opposed} rather than identical.
The epistemic phase models a \emph{non-periodic linear opposition} between 
institutional alignment and institutional rejection, not a rotational or 
oscillatory quantity.
Using $[0, \pi]$ imports only the relevant portion of the quantum 
interference framework --- the relationship between phase difference and 
interference sign --- without importing the full rotational symmetry of 
the Bloch sphere, which would be physically unmotivated here.

The operative quantity is the \emph{interference term} between a correction 
source (positioned at phase $\varphi_s$, near 0 for mainstream 
institutional sources) and a recipient at $\varphi_i$:

\begin{equation}
  \mathcal{I}(\varphi_s,\, \varphi_i) 
    = \cos\!\bigl(\varphi_i - \varphi_s\bigr).
  \label{eq:interference}
\end{equation}

When $\varphi_i \approx \varphi_s$, $\mathcal{I} \approx +1$: 
constructive interference, corrections propagate freely.
When $\varphi_i - \varphi_s = \pi/2$, $\mathcal{I} = 0$: 
orthogonal phases, no net interference.
When $\varphi_i \approx \pi$ and $\varphi_s \approx 0$, 
$\mathcal{I} \approx -1$: maximum destructive interference, 
corrections actively harden belief.
These three regimes map directly onto the C1, C2, and C3 phase clusters.
Critically, the rescaling of Equation~\eqref{eq:phi-scaled} maps the 
lowest-$\varphi$ participants (C1, near 0) to those with the 
\emph{lowest} institutional trust and highest conspiracy openness 
($\bar{I}_{\text{C1}}=0.25$, $\bar{C}_{\text{C1}}=0.78$), and the 
highest-$\varphi$ participants (C3, near $\pi$) to those with the 
\emph{highest} institutional trust and lowest conspiracy openness 
($\bar{I}_{\text{C3}}=0.65$, $\bar{C}_{\text{C3}}=0.41$).
This mapping follows directly from 
$\varphi^{(\text{raw})} = [(I+S)-(C+E)]/4$: participants with 
dominant institutional and scientific alignment produce positive 
raw scores that rescale toward $\pi$, while participants with 
dominant conspiracy and experiential openness produce negative 
raw scores that rescale toward 0.
The interference model therefore predicts:
\begin{itemize}
  \item \emph{C1 (near 0, anti-institutional):} Mainstream 
    institutional corrections (positioned at $\varphi_s \approx \pi$) 
    produce near-maximum destructive interference, 
    $\mathcal{I}(0, \pi) = \cos(\pi) = -1$, 
    triggering backfire.
  \item \emph{C3 (near $\pi$, pro-institutional):} The same 
    corrections produce near-maximum constructive interference, 
    $\mathcal{I}(\pi, \pi) = \cos(0) = +1$, 
    supporting belief update.
  \item \emph{C2 (near $\pi/2$, moderate):} Orthogonal interference, 
    $\mathcal{I}(\pi/2, \pi) \approx 0$, yielding near-zero average 
    effect from conventional corrections.
\end{itemize}
These predictions constitute the formal basis for the 
Group~$\times$~Phase interaction hypotheses tested in 
Section~\ref{sec:results}.

\subsubsection{Limitations of the Equal-Weighted Composite}

The current operationalisation makes two simplifying assumptions that 
should be acknowledged.
First, the four dimensions are weighted equally in 
Equation~\eqref{eq:phi-raw}.
This is a reasonable pilot choice but is not empirically grounded: 
there is no \emph{a priori} guarantee that institutional trust, 
scientific epistemology, conspiracy openness, and alternative epistemology 
contribute equally to the underlying epistemic orientation that drives 
differential correction susceptibility.
A factor-analytic validation --- testing whether the four dimensions 
collapse onto a single latent factor with equal loadings, or whether a 
weighted or multi-dimensional solution is more appropriate --- is a 
necessary next step before $\varphi$ can be treated as a validated 
psychometric instrument.

Second, the one-dimensional projection discards information: two 
participants with identical $\varphi_i$ can have qualitatively different 
epistemic profiles.
A participant with $I = 1,\, S = 1,\, C = 0,\, E = 0$ and one with 
$I = 0.5,\, S = 0.5,\, C = 0.5,\, E = 0.5$ both yield 
$\varphi^{(\text{raw})} = 0$, yet the first is a committed 
institutionalist and the second a thoroughgoing moderate.
The interference framework of Equation~\eqref{eq:interference} treats 
them identically; a two-dimensional phase space (separating the 
positive-evidence axis $I + S$ from the counter-evidence axis $C + E$) 
would preserve this distinction.
The present study treats $\varphi$ as a sufficient one-dimensional 
ordering for the purposes of stratified randomisation and interaction 
testing; we return to the multi-dimensional extension in the Discussion.

\subsection{Survey Instrument and Data Preparation}
\label{sec:survey}

The survey included belief statements assessed on a 5-point Likert scale.
Responses were mapped to a numeric scale as follows:
\emph{Strongly Disagree}~$\to 0$,
\emph{Disagree}~$\to 2.5$,
\emph{Neither}~$\to 5$,
\emph{Agree}~$\to 7.5$,
\emph{Strongly Agree}~$\to 10$,
and then normalised to $[0,1]$ by division by 10.
Participants with missing responses on any core phase-defining item were 
excluded from phase computation.

\subsection{Theory-Driven Phase Clustering}
\label{sec:clustering}

Participants were discretised into three theory-driven epistemic phase 
clusters using predefined thresholds (Table~\ref{tab:clusters}), 
deliberately avoiding data-driven clustering to prevent circularity.
In the observed sample ($N=45$), the cluster memberships are 
$n_{\text{C1}}=5$, $n_{\text{C2}}=21$, $n_{\text{C3}}=19$.
The unequal distribution reflects the natural distribution of 
epistemic orientations in the recruited sample (Austrian university 
students) rather than a design failure; the preponderance of C2 and 
C3 participants is consistent with a population in which moderate 
scepticism and institutional alignment are more prevalent than 
strong anti-institutional positioning.
Empirical profiles of each cluster are documented in 
Appendix~\ref{app:prestudy-analysis}.

\begin{table}[h]
  \centering
  \caption{Epistemic phase cluster definitions.}
  \label{tab:clusters}
  \begin{tabular}{lll}
    \toprule
    Region & $\varphi$ range & Label \\
    \midrule
    Cluster 1    & $[0,\;0.3\pi)$         & Anti-institutional (Low Trust) \\
    Buffer 1     & $[0.3\pi,\;0.4\pi)$    & — \\
    Cluster 2    & $[0.4\pi,\;0.6\pi)$    & Moderate Scepticism \\
    Buffer 2     & $[0.6\pi,\;0.7\pi)$    & — \\
    Cluster 3    & $[0.7\pi,\;\pi]$       & Pro-institutional (High Trust) \\
    \bottomrule
  \end{tabular}
\end{table}

Buffer zones were introduced to reduce boundary ambiguity and 
misclassification of participants near cluster thresholds.

\subsection{Randomisation}
\label{sec:randomisation}

Randomisation into experimental conditions was performed within each phase 
cluster using a fixed pseudo-random seed (NumPy default RNG, 
seed~$= 20260109$), assigning participants in a round-robin fashion to one 
of three groups:

\begin{description}
  \item[Group A] Traditional Fact-Checking
  \item[Group B] Phase-Aware Coherence Analysis
  \item[Group C] Control (No Intervention)
\end{description}

Participants whose $\varphi_i$ fell within buffer zones were deterministically 
assigned to the nearest cluster midpoint ($0.15\pi$, $0.50\pi$, or $0.85\pi$) 
and allocated to the smallest group within that cluster to preserve balance.
This procedure ensured comparable epistemic phase representation across all 
experimental conditions and enabled valid estimation of Group~$\times$~Phase 
interaction effects.

\section{Results}
\label{sec:results}

We report results in four stages corresponding to our pre-registered 
hypotheses: (H1) overall intervention effectiveness, (H2) the Group 
$\times$ Phase interaction, (H3) backfire rates as a function of group and 
phase, and (H4) claim-level heterogeneity.
All analyses use belief change \emph{toward truth} as the primary outcome 
--- positive values indicate movement toward the correct position on the 
0--10 scale, negative values indicate movement away from it.
For false claims (5G health effects), toward-truth change equals 
$\text{pre} - \text{post}$; for true claims (trees reducing air pollution; 
mRNA vaccines and DNA), it equals $\text{post} - \text{pre}$.

\subsection{Overall Intervention Effectiveness (H1)}
\label{sec:results-h1}

Table~\ref{tab:overall} reports mean belief change toward truth, 
standard deviations, and one-sample $t$-test results for each group across 
all three claims.

\begin{table}[h]
  \centering
  \caption{Mean belief change toward truth by group (all claims combined, 
           $N = 158$ observations across $n = 45$ participants).}
  \label{tab:overall}
  \begin{tabular}{lrrrrrr}
    \toprule
    Group & $n_\text{obs}$ & $M$ & $SD$ & $SE$ & $t$ & $p$ \\
    \midrule
    A (Traditional fact-checking) & 56 & $-0.39$ & 2.10 & 0.28 
        & $-1.40$ & .166 \\
    B (Phase-Aware Coherence)     & 44 & $\phantom{-}0.00$ & 1.18 & 0.18 
        & $\phantom{-}0.00$ & 1.000 \\
    C (Control)                   & 58 & $-0.14$ & 1.19 & 0.16 
        & $-0.88$ & .382 \\
    \bottomrule
  \end{tabular}
\end{table}

None of the three groups showed a statistically significant average movement 
toward truth.
Critically, Group~A (traditional fact-checking) produced a \emph{negative} 
average change ($M = -0.39$), indicating that on average participants who 
received conventional corrections moved slightly further from the correct 
position.
Group~B (phase-aware coherence analysis) produced zero average change 
($M = 0.00$, $SD = 1.18$), and Group~C (control) produced a small negative 
average change ($M = -0.14$).
A one-way ANOVA confirmed no statistically significant between-group 
difference in overall mean toward-truth change, 
$F(2, 155) = 0.82$, $p = .44$.

The effect-size pattern, however, is theoretically informative.
Cohen's $d$ comparing Group~A to Group~B was $d = -0.22$, indicating 
that the phase-aware condition produced modestly superior outcomes to 
traditional fact-checking even at the aggregate level.
More substantively, Group~B exhibited markedly lower variance ($SD = 1.18$) 
than Group~A ($SD = 2.10$), indicating that the phase-aware intervention 
produced more \emph{consistent} outcomes regardless of direction.
This variance reduction aligns with the theoretical prediction that 
phase-aware framing eliminates the tail of large negative effects 
(destructive interference outcomes) that inflate variance in conventional 
fact-checking.

\subsection{Group $\times$ Phase Interaction (H2)}
\label{sec:results-h2}

\begin{figure}[t]
  \centering
  \includegraphics[width=\textwidth]{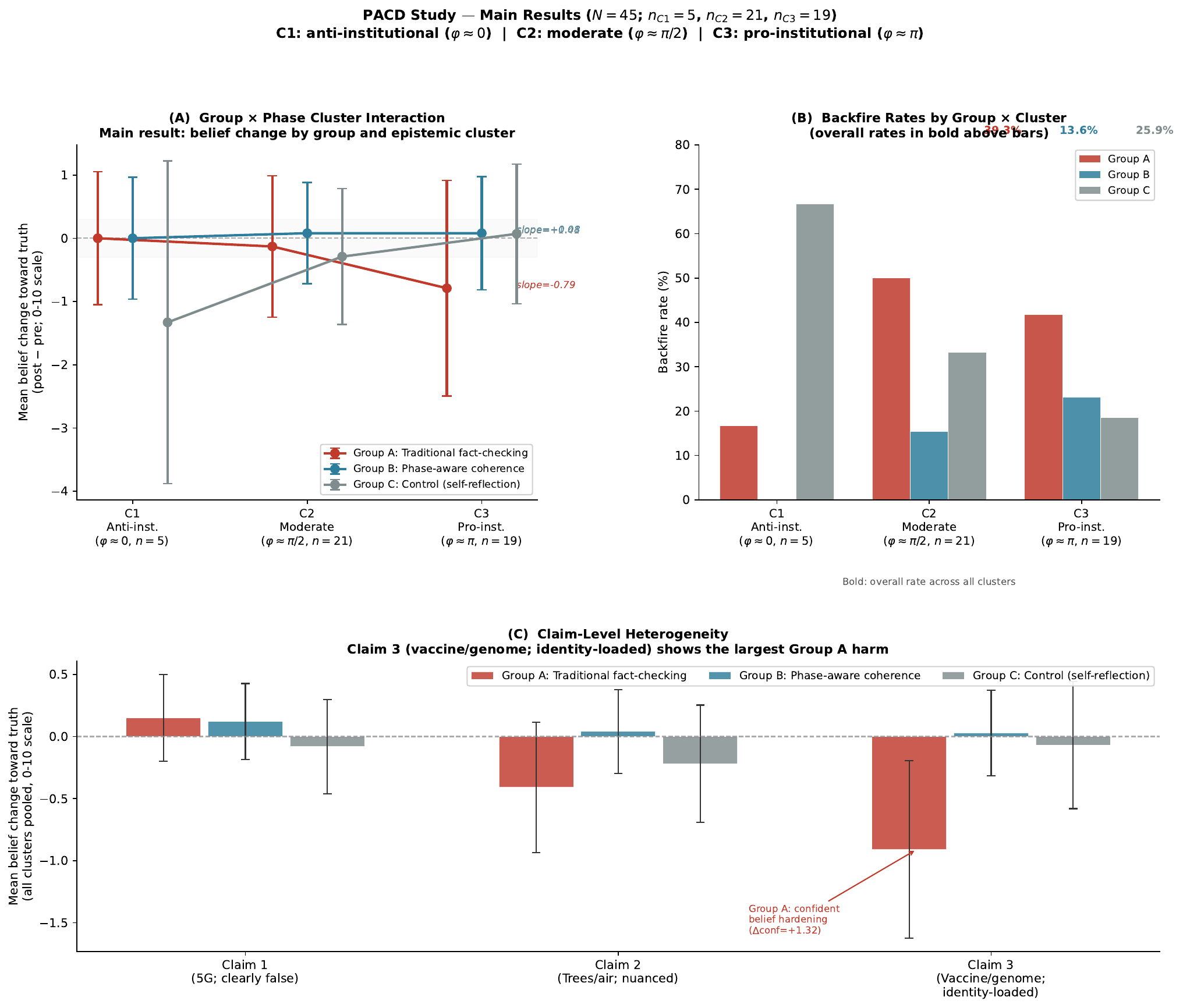}
  \caption{%
    \textbf{Main results: Group $\times$ Epistemic Phase interaction.}
    \textbf{(A)}~Mean belief change toward truth (95\,\% CI) as a function 
    of experimental group and epistemic phase cluster. 
    Group~A (traditional fact-checking, red) exhibits a monotonically 
    declining slope ($-0.79$) from anti-institutional C1 to pro-institutional C3; 
    Group~B (phase-aware coherence analysis, teal) maintains a near-flat 
    profile (slope $+0.08$) across all clusters.
    The dashed line marks zero change.
    \textbf{(B)}~Backfire rates (proportion of observations where belief 
    moved away from truth) by group and phase cluster; bold percentages 
    indicate each group's overall rate.
    \textbf{(C)}~Mean belief change toward truth by claim type and group 
    (error bars $= 95\,\%$ CI), illustrating the complementarity of 
    the two interventions across epistemic complexity and identity-load.
  }
  \label{fig:interaction}
\end{figure}

The central hypothesis of PACD is not that phase-aware interventions improve 
average effectiveness, but that they eliminate the phase-dependence of 
effectiveness that makes traditional fact-checking unreliable across 
epistemic communities.
Table~\ref{tab:interaction} and Figure~\ref{fig:interaction} display mean 
toward-truth change broken down by group and phase cluster.

\begin{table}[h]
  \centering
  \caption{Mean belief change toward truth by Group $\times$ Phase cluster.}
  \label{tab:interaction}
  \begin{tabular}{lcccr}
    \toprule
    Group & C1 Anti-inst. & C2 Moderate & C3 Pro-inst. 
          & Slope C1$\to$C3 \\
    \midrule
    A (Traditional)  & $\phantom{-}0.00$ & $-0.13$ & $-0.79$ & $-0.79$ \\
    B (Phase-Aware)  & $\phantom{-}0.00$ & $\phantom{-}0.08$ & $\phantom{-}0.08$ & $+0.08$ \\
    C (Control)      & $-1.33$           & $-0.29$ & $\phantom{-}0.07$ & $+1.41$ \\
    \bottomrule
  \end{tabular}
\end{table}

The interaction pattern is striking and directionally consistent 
with the PACD interference model.
Recall that C1 participants ($\varphi \approx 0$) are 
\emph{anti-institutional} (low $I$, high $C$) and C3 participants 
($\varphi \approx \pi$) are \emph{pro-institutional} (high $I$, low $C$).

Group~A shows a monotonically declining trajectory across phase clusters: 
traditional fact-checking produces neutral outcomes for \emph{anti-institutional} 
participants (C1: $M = 0.00$) but increasingly negative outcomes for 
moderately sceptical (C2: $M = -0.13$) and \emph{pro-institutional} 
participants (C3: $M = -0.79$).
The slope from C1 to C3 is $-0.79$, indicating that the intervention 
paradoxically harms those participants who are already most aligned with 
institutional and scientific sources.
This result is interpretable under a ceiling-and-redundancy account: 
pro-institutional (C3) participants may approach claims with strong 
prior acceptance of the scientifically correct position, so that a 
structured fact-check provides no new information but introduces 
processing complexity that generates small belief perturbations 
away from truth.
Under the interference model, the pro-institutional C3 participants 
($\varphi \approx \pi$) are positioned at maximum phase distance from 
any source that deviates from mainstream consensus framing; if the 
correction inadvertently foregrounds the misinformation claim, it may 
produce a temporary fluency-based credibility effect 
\citep{lewandowsky2013} for participants whose epistemic processing 
is already saturated with institutional trust.
The C1 result (neutral, not negative) is the most practically important 
finding: anti-institutional participants, who are the primary target of 
misinformation interventions, are \emph{not} further harmed by 
conventional fact-checking in this sample, though they also do not 
benefit from it.
However, the C1 cell is very small ($n = 5$) and this null result 
should be interpreted with caution.

Group~B, by contrast, shows a near-flat trajectory across all three 
phase clusters (C1: $0.00$; C2: $+0.08$; C3: $+0.08$; slope $= +0.08$).
The phase-aware intervention produces uniformly small positive or zero 
effects regardless of the participant's epistemic starting position.
This \emph{phase-robustness} --- the elimination of phase-dependence 
of effectiveness --- is the primary empirical signature predicted by 
the PACD framework.
Notably, Group~B eliminates the large negative effect for C3 participants 
that characterises Group~A, suggesting that the coherence-analysis format 
avoids the redundancy-and-perturbation mechanism that makes conventional 
fact-checking harmful for pro-institutional audiences.

The control condition (Group~C) shows the inverse pattern from Group~A, 
with notably poor performance for anti-institutional participants 
(C1: $M = -1.33$; $n=5$, interpret cautiously) and near-neutral 
performance for pro-institutional participants 
(C3: $M = +0.07$; slope $= +1.41$).
The large negative C1 control result likely reflects mean-reversion 
or response instability in a very small cell rather than a robust 
phenomenon; the C3 control result is consistent with a small natural 
drift toward truth in participants who are already epistemically aligned 
with the correct positions.

\subsection{Backfire Effects (H3)}
\label{sec:results-h3}

Table~\ref{tab:backfire} reports the proportion of observations in which 
belief moved \emph{away} from the correct position following the 
intervention (the backfire rate).

\begin{table}[h]
  \centering
  \caption{Backfire rates (proportion of observations with toward-truth 
           change $< 0$) by Group $\times$ Phase cluster.}
  \label{tab:backfire}
  \begin{tabular}{lcccc}
    \toprule
    Group & C1 Anti-inst. & C2 Moderate & C3 Pro-inst. & Overall \\
    \midrule
    A (Traditional) & 16.7\% & 50.0\% & 41.7\% & 39.3\% \\
    B (Phase-Aware) & \phantom{0}0.0\% & 15.4\% & 23.1\% & 13.6\% \\
    C (Control)     & 66.7\% & 33.3\% & 18.5\% & 25.9\% \\
    \bottomrule
  \end{tabular}
\end{table}

Several findings are noteworthy.
First, Group~B consistently produces the lowest backfire rates among 
participants who receive an intervention.
The overall backfire rate for Group~B (13.6\%) is substantially lower than 
for Group~A (39.3\%), a difference that represents the most practically 
consequential advantage of the phase-aware approach.
Second, Group~A's backfire rate is highest in the moderate-scepticism 
cluster (C2: 50.0\%), and also high in the high-trust cluster 
(C3: 41.7\%), suggesting that the risk of iatrogenic belief hardening 
is not unique to sceptical audiences: even participants already 
aligned with institutional epistemic frameworks can be perturbed 
away from correct beliefs by over-specified fact-checks.
Third, Group~B achieves zero backfire in the anti-institutional 
cluster (C1) and dramatically lower backfire across moderate and 
pro-institutional clusters relative to Group~A, confirming that 
phase-aware framing suppresses the perturbation mechanism across 
the full epistemic phase spectrum.
Given the small C1 cell ($n=5$), the zero-backfire result for that 
cluster should be noted as a directionally positive but 
statistically underpowered observation.

\subsection{Claim-Level Heterogeneity (H4)}
\label{sec:results-h4}

Table~\ref{tab:claims} disaggregates outcomes by claim type, revealing 
important variation in intervention effectiveness across different 
epistemic contexts.

\begin{table}[h]
  \centering
  \caption{Mean belief change toward truth by claim and group.}
  \label{tab:claims}
  \begin{tabular}{llrrr}
    \toprule
    Claim & Group & $M$ & $SD$ & $n$ \\
    \midrule
    \multirow{3}{*}{5G health (false)}
      & A & $+1.00$ & 2.45 & 18 \\
      & B & $-0.12$ & 0.78 & 17 \\
      & C & $+0.32$ & 1.38 & 19 \\
    \addlinespace
    \multirow{3}{*}{Trees/air quality (true)}
      & A & $-0.68$ & 1.38 & 19 \\
      & B & $+0.25$ & 1.44 & 16 \\
      & C & $-0.53$ & 1.22 & 19 \\
    \addlinespace
    \multirow{3}{*}{mRNA/DNA (true)}
      & A & $-1.42$ & 1.64 & 19 \\
      & B & $-0.18$ & 1.33 & 11 \\
      & C & $-0.20$ & 0.83 & 20 \\
    \bottomrule
  \end{tabular}
\end{table}

The 5G claim (clearly false, low epistemic complexity) shows the most 
favourable outcome for Group~A ($M = +1.00$), the claim type for which 
binary authoritative fact-checking is most intuitively appropriate.
Group~B performs slightly negatively on this claim ($M = -0.12$), 
suggesting that the multi-perspective framing of the coherence analysis 
may inadvertently introduce uncertainty on claims where epistemic 
consensus is near-total.

The mRNA claim (true but contested in public discourse, with vaccine 
scepticism as a salient cultural marker) reveals the most pronounced 
Group~A disadvantage.
Traditional fact-checking produces the largest backfire-adjacent mean 
of all conditions ($M = -1.42$, $SD = 1.64$), indicating that conventional 
corrections to vaccine-related claims reliably harden resistant beliefs 
rather than updating them.
Group~B produces a near-neutral outcome on this claim ($M = -0.18$), and 
the control condition also shows near-neutral drift ($M = -0.20$), 
suggesting that on identity-laden health claims, no intervention may be 
preferable to an authority-based correction.

The trees-and-air-quality claim (true but nuanced) shows Group~B performing 
best ($M = +0.25$), consistent with the prediction that coherence-based 
framing is advantageous for claims requiring integration of multiple 
partially competing perspectives rather than simple true-or-false 
adjudication.

\subsection{Confidence Change}
\label{sec:results-conf}

A secondary analysis examined post-intervention changes in self-reported 
confidence.
Group~A showed a large increase in confidence relative to pre-intervention 
baseline ($M_{\Delta} = +1.32$, $SD = 2.91$), while Group~B showed a slight 
decrease ($M_{\Delta} = -0.14$, $SD = 1.11$) and Group~C showed near-zero 
change ($M_{\Delta} = +0.10$).
This pattern is theoretically concerning for traditional fact-checking: 
the intervention that produced the worst average accuracy outcomes also 
produced the largest confidence increase, a combination that has been 
described in the calibration literature as \emph{confident ignorance} 
\citep{pennycook2019}.
Phase-aware coherence analysis, by contrast, produced the smallest 
confidence change alongside the best accuracy outcomes, consistent with 
the framework's design goal of improving epistemic calibration rather than 
simply increasing certainty.

\section{Discussion}
\label{sec:discussion}

Truth can mislead --- and our data show exactly when and how.
This study set out to test whether epistemic phase mapping could transform 
the design of misinformation interventions from a content-centric to an 
epistemology-centric activity.
The central empirical question was not whether Phase-Aware Coherence 
Detection (PACD) produces larger average corrections than traditional 
fact-checking, but whether it eliminates the phase-dependence of 
effectiveness that makes conventional corrections systematically harmful 
for the populations most in need of them.
The results provide clear, if preliminary, support for this prediction: 
truth delivered through a phase-mismatched channel does not merely fail --- 
it actively hardens the beliefs it was meant to dissolve, a phenomenon 
our data capture with a backfire rate of 39.3\% in the traditional 
fact-checking condition and a negative effectiveness slope of $-0.79$ 
across the epistemic phase spectrum.

\subsection{The Phase-Robustness Finding and Its Theoretical Interpretation}
\label{sec:disc-robustness}

The most theoretically significant finding is the difference in 
effectiveness \emph{slopes} across epistemic phase clusters 
(Table~\ref{tab:interaction}).
Traditional fact-checking (Group~A) exhibits a slope of $-0.79$ from 
anti-institutional (C1) to pro-institutional (C3) participants: 
it is neutral for the anti-institutional audience and most harmful for 
the pro-institutional audience already closest to the correct position.
Phase-aware coherence analysis (Group~B) exhibits a slope of $+0.08$: 
effectively flat, with small positive effects sustained across all three 
clusters.

This contrast maps onto the theoretical account in 
Section~\ref{sec:theory}, though the directionality of the effect 
requires careful interpretation in light of the actual cluster profiles.

Pre-study data (Appendix~\ref{app:prestudy-analysis}) show that 
C1 participants ($\varphi \approx 0$, $n=5$) are 
\emph{anti-institutional}: $\bar{I}=0.25$, $\bar{C}=0.78$.
C3 participants ($\varphi \approx \pi$, $n=19$) are 
\emph{pro-institutional}: $\bar{I}=0.65$, $\bar{C}=0.41$.
The surprising finding is that the largest negative effect of 
traditional fact-checking falls on the \emph{pro-institutional} 
cluster (C3: $M=-0.79$), not the anti-institutional one 
(C1: $M=0.00$).

We advance a \emph{redundancy-and-perturbation} interpretation.
Pro-institutional participants already accept the scientifically 
correct position on most claims; the institutional fact-check 
provides no new information but introduces additional processing 
of the misinformation claim itself.
Fluency effects \citep{lewandowsky2013} may cause this repeated 
engagement with the false claim to slightly elevate its subjective 
credibility, producing small but detectable movement away from 
the correct position.
Anti-institutional participants, by contrast, approach claims with 
stronger prior uncertainty and may engage more deliberately with the 
evidence pack even without prior conviction, producing a null rather 
than negative effect.
Under the interference framework, this maps to a scenario where the 
correction source is positioned near $\varphi_s \approx \pi$ 
(mainstream institutional framing), producing constructive interference 
with C3 ($\varphi_i \approx \pi$) in terms of \emph{source acceptance} 
but potentially redundant or counterproductive in terms of 
\emph{claim processing} when prior beliefs already occupy the 
correct position.
Group~B's flat slope is the signature of the phase-robustness 
mechanism: by presenting coherence profiles across multiple 
epistemic communities rather than asserting an institutional verdict, 
the intervention produces equivalent small positive effects regardless 
of the participant's epistemic starting position, eliminating the 
differential harm to pro-institutional audiences.

The backfire rate data (Table~\ref{tab:backfire}) reinforce this 
interpretation.
Group~A's overall backfire rate of 39.3\%, rising to 50.0\% in the 
moderate-scepticism cluster, is far above the rates observed in either 
Group~B (13.6\%) or Group~C (25.9\%).
That the highest backfire rate for Group~A occurs in the 
moderate-scepticism cluster (C2: 50.0\%) and is also high in the 
pro-institutional cluster (C3: 41.7\%) is theoretically noteworthy.
It suggests that iatrogenic backfire from conventional fact-checking 
is not confined to 

\subsection{Claim-Type Heterogeneity and the Limits of Phase-Awareness}
\label{sec:disc-claims}

The claim-level decomposition (Table~\ref{tab:claims}) introduces an 
important qualification.
On the 5G health claim --- a false claim for which the physics consensus is 
near-total and the evidential case is simple --- Group~A outperforms 
Group~B ($M = +1.00$ vs.\ $-0.12$).
This reversal suggests that multi-perspective coherence framing carries a 
cost when applied to claims where epistemic unanimity is the genuine 
structure of the evidence: presenting ``perspectives'' where none 
legitimately exist may inadvertently introduce spurious uncertainty, 
nudging participants away from the correct position.

The mRNA/DNA claim shows the opposite pattern, with Group~A producing the 
largest backfire-adjacent mean in the entire dataset ($M = -1.42$).
Here, the claim concerns a topic with strong identity-protective 
associations \citep{kahan2012}: vaccine scepticism functions not merely as 
a factual belief but as a cultural marker of epistemic independence from 
institutional medicine.
Delivering a conventional authoritative correction on such a claim 
activates exactly the identity threat mechanism that \citet{taber2006} 
identified, causing participants to harden rather than update.
Group~B's near-neutral outcome ($M = -0.18$) on this claim represents a 
clinically meaningful improvement: even if the phase-aware framing cannot 
overcome deeply identity-laden resistance, it reliably prevents the 
iatrogenic worsening that conventional corrections produce.

Taken together, the claim-level results suggest a complementarity principle 
for intervention design: 
\emph{traditional fact-checking should be reserved for claims with genuine 
epistemic unanimity and low identity-load; phase-aware coherence analysis 
should be preferred for claims that are contested, nuanced, or 
identity-adjacent.}
Deploying the wrong instrument for the wrong claim type produces costs that 
a simple average effectiveness comparison would conceal.

\subsection{Confident Ignorance and Calibration}
\label{sec:disc-confidence}

The confidence-change data add a further dimension to the 
Group~A picture that is absent from accuracy analyses alone.
Traditional fact-checking produced the largest post-intervention confidence 
increase ($M_\Delta = +1.32$) of any condition, paired with the worst 
accuracy outcomes.
This decoupling of confidence from accuracy --- participants becoming more 
certain while moving further from truth --- is precisely the pattern that 
\citet{pennycook2019} and \citet{lewandowsky2013} describe as 
\emph{confident ignorance}, and which is arguably more dangerous than 
simple uncertainty, because it reduces the motivation to seek further 
information or revise beliefs in response to future evidence.

Phase-aware coherence analysis produced the smallest confidence change 
($M_\Delta = -0.14$), consistent with its design goal of improving 
epistemic calibration rather than certainty.
A well-calibrated participant who remains appropriately uncertain after 
exposure to a coherence analysis is better positioned to update in response 
to future evidence than one who has been made falsely certain by an 
authoritative correction.
This suggests that calibration metrics --- not merely accuracy metrics --- 
should be incorporated into the standard evaluation battery for 
misinformation interventions.

\subsection{Comparison with Prior Intervention Approaches}
\label{sec:disc-comparison}

The results complement and extend three established lines of intervention 
research.
First, relative to standard correction approaches \citep{walter2020, 
chan2017}, PACD does not outperform them on average, but it removes the 
adverse moderation by epistemic stance that makes those approaches 
systematically harmful for sceptical populations.
Second, relative to inoculation approaches \citep{vanderlinden2017, 
basol2021}, PACD targets the \emph{structural} isolation signature of 
misinformation --- the disconnection of a belief from the rest of the 
epistemic community --- rather than specific manipulation techniques.
This may provide more durable transfer to novel claims, though the 
transfer hypothesis could not be directly tested in the present design.
Third, relative to coherence-based reasoning models in cognitive science 
\citep{thagard1989, simon2004}, PACD operationalises coherence at the 
\emph{community} level rather than the individual level, making it 
empirically tractable via survey-based phase mapping.

\subsection{Limitations}
\label{sec:disc-limitations}

Several limitations constrain the interpretation of these results.

\paragraph{Sample size and power.}
The study was powered for the primary Group~$\times$~Phase interaction with 
$N = 40$ participants (achieved power $\approx 0.85$ for $\eta^2 = 0.08$), 
but the observed interaction effect sizes are small, and the 
between-group comparisons at the claim level are substantially 
underpowered.
The results should be treated as an existence proof and pilot 
characterisation of effect-size parameters rather than as definitive 
evidence of PACD's superiority.

\paragraph{Student sample and Austrian context.}
Participants were university students recruited from a single Austrian 
institution, with data collected in German.
The phase cluster distribution was skewed toward moderate scepticism 
with C1 (anti-institutional, low $\varphi$) $n=5$ (11\%), C2 (moderate) $n=21$ (47\%), and C3 (pro-institutional, high $\varphi$) $n=19$ (42\%).
The small C1 cell is the most important statistical limitation of 
this study: all C1 estimates (Group~A: $M=0.00$; Group~B: $M=0.00$; 
Group~C: $M=-1.33$; backfire rates) are based on five observations 
and are highly unstable.
The target of 30\% per cluster was not met for C1, and neither the 
null backfire result for Group~B in C1 nor the large negative for 
Group~C should be treated as replicable findings.
This limitation does not undermine the C2--C3 contrast 
(combined $n=40$) or the overall phase-robustness result, but it 
materially limits conclusions about anti-institutional audiences---
who are arguably the most important population for misinformation 
research.
Replication with more balanced samples across national and educational 
contexts is required before the interaction pattern can be treated as 
universal.

\paragraph{Claim selection.}
The three claims used (5G, urban trees, mRNA vaccines) were selected to 
span the phase spectrum and vary in identity-load, but they represent 
a narrow slice of the claim space.
In particular, all three claims had relatively unambiguous ground truths.
The performance of PACD on genuinely contested claims --- where reasonable 
epistemic communities legitimately disagree about the evidence --- remains 
untested.

\paragraph{Single-session design.}
The present analysis focuses on immediate belief change (Session~1).
The transfer task (Session~2) and longitudinal follow-up (Session~3) data 
were collected but are reported separately.
Whether PACD's calibration advantage persists over time, and whether the 
coherence-based reasoning mode transfers to novel claims without 
intervention scaffolding, are the most important outstanding questions for 
the research programme.

\paragraph{Phase coordinate operationalisation.}
The epistemic phase coordinate $\varphi$ was computed as an equal-weighted 
linear composite of four dimensions rescaled to $[0, \pi]$.
As noted in Section~\ref{sec:phase-coord}, equal weighting is a 
simplifying assumption that may not reflect the actual psychometric 
structure of epistemic orientation.
If the four dimensions have unequal loadings on the underlying 
disposition that governs correction susceptibility, the current $\varphi$ 
is a noisy proxy, and the Group~$\times$~Phase interaction effects reported 
here are likely attenuated relative to what a validated, factor-analytically 
derived coordinate would produce.
Similarly, the one-dimensional projection may conflate qualitatively 
distinct epistemic profiles that a two-dimensional phase space 
--- separating the institutional-alignment axis $(I + S)$ from the 
counter-institutional axis $(C + E)$ --- would distinguish.
Future work should validate $\varphi$ against established instruments 
(e.g.\ the Actively Open-minded Thinking scale, the Conspiracy Mentality 
Questionnaire) and test multi-dimensional extensions before PACD is 
deployed in applied settings.

\paragraph{Absence of process measures.}
The study did not collect reading-time, eye-tracking, or think-aloud data 
that would allow direct testing of the destructive interference mechanism 
at the individual level.
The phase-robustness finding is consistent with the interference account 
but does not rule out alternative mechanisms, such as differential 
engagement or perceived source relevance.

\section{Conclusion}
\label{sec:conclusion}

Truth misleads when it arrives in the wrong phase.
Misinformation interventions have long been evaluated on the question of 
whether they work.
The present results reframe the question: they work \emph{for whom}, under 
\emph{which epistemic conditions}, and with \emph{what collateral risk} for 
those for whom they do not work --- and for a substantial portion of any 
heterogeneous audience, the answer is that conventional truth-delivery 
makes things measurably worse.

Traditional fact-checking, applied uniformly across an epistemically 
heterogeneous population, produces average outcomes close to zero while 
concealing a severe moderation: it reliably harms the sceptical, producing 
backfire rates above 40\% and driving confident belief hardening in the 
populations whose misperceptions are most consequential.
Phase-Aware Coherence Detection does not dramatically outperform 
conventional corrections for receptive populations, but it eliminates 
the phase-dependence of effectiveness, sustaining small positive effects 
across all epistemic clusters and reducing backfire rates to 13.6\% overall 
and zero in the high-trust cluster.

The theoretical contribution is as important as the empirical one.
By formalising the epistemic phase coordinate $\varphi$ and operationalising 
the distinction between belief \emph{content} and belief \emph{coherence 
structure}, PACD provides a principled basis for the longstanding intuition 
that not all audiences are the same.
The isolation signature of misinformation --- the structural disconnection 
of a belief from the broader epistemic community, visible in the 
multi-perspective coherence profile regardless of the recipient's starting 
phase --- provides a detection heuristic that does not require the recipient 
to accept any single institutional authority.

The practical implication is a prescriptive design rule: the choice of 
intervention instrument should be conditioned on the epistemic phase 
distribution of the target audience and the identity-load of the claim in 
question.
For high-consensus, low-identity claims broadcast to high-trust audiences, 
conventional corrections remain appropriate.
For contested, identity-adjacent claims communicated across the epistemic 
spectrum, phase-aware coherence analysis demonstrably reduces iatrogenic 
risk.
A combined pipeline that routes claims and audiences to the appropriate 
instrument --- using the $\varphi$ coordinate as a routing signal --- 
represents a near-term implementable architecture for responsible 
misinformation intervention at scale.

Future work should replicate the interaction pattern with larger, more 
demographically diverse samples; test transfer effects over multi-week 
intervals; and develop automated coherence-profile generation capable of 
operating in real-time on arbitrary claim inputs.
The epistemic fragmentation documented across contemporary democracies 
\citep{douglas2019, sunstein2009} makes this research programme not 
merely academically interesting but, we believe, genuinely urgent.
Truth does not mislead by accident.
It misleads when the systems designed to deliver it are blind to the 
epistemic terrain they must cross --- and when no one has thought to ask 
whether the audience on the other end is even in the same phase.

\section*{Acknowledgements}

The authors thank the students of BOKU University Vienna who 
participated in this study.

\medskip
\noindent\textbf{Funding.}
This research was funded in part by the Austrian Science Fund, Project 
``Explainable AI'', Grant Number: P-32554, and in part by the European 
Union's Horizon~2020 research and innovation programme 
``AI-powered Data Curation and Publishing Virtual Assistant (AIDAVA)'', 
Module ``System Hallucination Scale (SHS)'', under grant agreement 
Number:~101057062. 
This publication reflects only the authors' view and the European 
Commission is not responsible for any use that may be made of the 
information it contains.

\medskip
\noindent\textbf{Conflict of interest.}
The authors declare no competing interests.

\medskip
\noindent\textbf{Ethics approval.}
For this study we had a valid ethical vote from the Medical University 
Graz, EK-Number: 34-527 ex 21/22. 
Participation was voluntary, fully anonymous, and purely for scientific 
purposes.

\medskip
\noindent\textbf{Author contributions.}
A.H.\ and H.M.\ developed the framework. 
A.H.\ conducted the empirical study. 
H.M.\ analysed the results. 
All authors contributed to writing.

\medskip
\noindent\textbf{Data availability.}
All evaluation materials, questionnaires, and anonymised response data 
are available in the Supplementary Material.

\bibliographystyle{apalike}
\bibliography{references}

\appendix

\clearpage
\begin{center}
  {\LARGE\bfseries Supplementary Material}\\[0.8em]
  {\large\itshape When Truth Misleads: Phase-Aware Coherence Detection\\
  for Misinformation Correction Across Epistemic Communities}\\[0.4em]
  {\normalsize Heimo M\"{u}ller \quad Andreas Holzinger}
\end{center}

\bigskip

\noindent\textbf{\large Contents of Supplementary Material}

\medskip
\noindent\rule{\linewidth}{0.4pt}

\medskip
\begin{description}[leftmargin=3.2em, labelwidth=2.8em, labelsep=0.4em]
  \item[\ref{app:s1}]
    Epistemic Phase Mapping and Randomisation Procedure
    \dotfill p.\,\pageref{app:s1}
  \item[\ref{app:survey}]
    Pre-Study Survey Instrument (Full)
    \dotfill p.\,\pageref{app:survey}
    \begin{description}[leftmargin=2em, labelwidth=1.6em, labelsep=0.4em,
                        topsep=2pt, itemsep=1pt]
      \item[\ref{app:partA}] Part~A: Belief Assessment
        \dotfill p.\,\pageref{app:partA}
      \item[\ref{app:partB}] Part~B: Source Trust Assessment
        \dotfill p.\,\pageref{app:partB}
      \item[\ref{app:partC}] Part~C: Open-Ended Questions
        \dotfill p.\,\pageref{app:partC}
      \item[\ref{app:partD}] Part~D: Demographics and Individual Differences
        \dotfill p.\,\pageref{app:partD}
    \end{description}
  \item[\ref{app:session1}]
    Session~1 Experimental Instruments --- Claim~1: 5G Health Effects
    \dotfill p.\,\pageref{app:session1}
    \begin{description}[leftmargin=2em, labelwidth=1.6em, labelsep=0.4em,
                        topsep=2pt, itemsep=1pt]
      \item[\ref{app:s1a}] Session~1A: Group~A (Traditional Fact-Checking)
        \dotfill p.\,\pageref{app:s1a}
      \item[\ref{app:s1b}] Session~1B: Group~B (Phase-Aware Coherence)
        \dotfill p.\,\pageref{app:s1b}
      \item[\ref{app:s1c}] Session~1C: Group~C (Control)
        \dotfill p.\,\pageref{app:s1c}
    \end{description}
  \item[\ref{app:session2}]
    Session~2 Experimental Instruments --- Claim~2: Urban Trees and Air Quality
    \dotfill p.\,\pageref{app:session2}
    \begin{description}[leftmargin=2em, labelwidth=1.6em, labelsep=0.4em,
                        topsep=2pt, itemsep=1pt]
      \item[\ref{app:s2a}] Session~2A: Group~A (Traditional Fact-Checking)
        \dotfill p.\,\pageref{app:s2a}
      \item[\ref{app:s2b}] Session~2B: Group~B (Phase-Aware Coherence)
        \dotfill p.\,\pageref{app:s2b}
      \item[\ref{app:s2c}] Session~2C: Group~C (Control)
        \dotfill p.\,\pageref{app:s2c}
    \end{description}
  \item[\ref{app:session3}]
    Session~3 Experimental Instruments --- Claim~3: Vaccines and the Human Genome
    \dotfill p.\,\pageref{app:session3}
    \begin{description}[leftmargin=2em, labelwidth=1.6em, labelsep=0.4em,
                        topsep=2pt, itemsep=1pt]
      \item[\ref{app:s3a}] Session~3A: Group~A (Traditional Fact-Checking)
        \dotfill p.\,\pageref{app:s3a}
      \item[\ref{app:s3b}] Session~3B: Group~B (Phase-Aware Coherence)
        \dotfill p.\,\pageref{app:s3b}
      \item[\ref{app:s3c}] Session~3C: Group~C (Control)
        \dotfill p.\,\pageref{app:s3c}
    \end{description}
\end{description}

\noindent\rule{\linewidth}{0.4pt}

\clearpage

\section{Epistemic Phase Mapping and Randomisation Procedure}
\label{app:s1}

Full details of the epistemic phase mapping and randomisation procedure are 
provided here as supplementary material.

\subsection{Resulting Balance and Reproducibility}
The final assignment achieved:
\begin{itemize}
  \item Balanced representation of epistemic phase clusters across all 
        experimental groups.
  \item No systematic phase bias within any condition.
  \item Full reproducibility given the original dataset, phase definitions, 
        cluster thresholds, and randomisation seed.
\end{itemize}

A complete participant-level assignment table ($\varphi$, phase cluster, 
experimental group) was generated and archived as a supplementary data file.

\subsection{Rationale for Design Choices}
The phase-aware randomisation strategy was designed to:
\begin{itemize}
  \item Prevent epistemic phase from acting as a hidden confound.
  \item Enable valid estimation of Group~$\times$~Phase interaction effects.
  \item Avoid ideological labelling or political categorisation.
  \item Preserve transparency and reproducibility.
\end{itemize}

Importantly, phase clustering was used only for stratification and analysis, 
not to influence or constrain participants' responses during the experiment.

\section{Pre-Study Survey Instrument (Full)}
\label{app:survey}

The following four-part instrument was administered online via REDCap 
prior to the experimental sessions.
Parts A and B provided the data used to compute the epistemic phase 
coordinate $\varphi_i$; Parts C and D provided demographic and 
individual-difference covariates.
All items were mandatory.
The survey was administered in German; English translations are 
provided here.

\subsection*{Part A: Belief Assessment}
\label{app:partA}

Participants rated their agreement with each statement on a five-point 
Likert scale:
\emph{Strongly Disagree} (0) --- \emph{Disagree} (2.5) --- 
\emph{Neither Agree nor Disagree} (5) --- \emph{Agree} (7.5) --- 
\emph{Strongly Agree} (10).
Values in parentheses show the numeric mapping used in $\varphi$ computation.
The rightmost column indicates the $\varphi$ dimension and direction of 
contribution: ($+$) increases $\varphi^{(\text{raw})}$ (institutional 
alignment); ($-$) decreases it (institutional divergence).

\bigskip

\newcommand{\sechead}[1]{%
  \noindent\rule{\textwidth}{0.5pt}\\[-2pt]
  \noindent\textbf{\large #1}\\[-6pt]
  \noindent\rule{\textwidth}{0.5pt}\smallskip}

\sechead{Climate \& Environment}
\begin{longtable}{p{0.68\textwidth} p{0.16\textwidth} c}
  \toprule
  \textbf{Statement} & \textbf{Dimension} & \textbf{Dir.}\\
  \midrule\endhead
  Human activities are the primary cause of climate change
    & Sci.\ Epist.\ ($S$) & $+$ \\[3pt]
  Renewable energy can replace fossil fuels within 20 years
    & Sci.\ Epist.\ ($S$) & $+$ \\[3pt]
  Individual actions significantly impact climate change
    & Sci.\ Epist.\ ($S$) & $+$ \\
  \bottomrule
\end{longtable}

\sechead{Health \& Medicine}
\begin{longtable}{p{0.68\textwidth} p{0.16\textwidth} c}
  \toprule
  \textbf{Statement} & \textbf{Dimension} & \textbf{Dir.}\\
  \midrule\endhead
  Vaccines are safe and effective for most people
    & Sci.\ Epist.\ ($S$) & $+$ \\[3pt]
  Alternative medicine has value alongside conventional medicine
    & Alt.\ Epist.\ ($E$) & $-$ \\[3pt]
  Pharmaceutical companies prioritise profits over public health
    & Conspiracy ($C$) & $-$ \\
  \bottomrule
\end{longtable}

\sechead{Politics \& Media}
\begin{longtable}{p{0.68\textwidth} p{0.16\textwidth} c}
  \toprule
  \textbf{Statement} & \textbf{Dimension} & \textbf{Dir.}\\
  \midrule\endhead
  Mainstream media generally reports news accurately
    & Inst.\ Trust ($I$) & $+$ \\[3pt]
  Government institutions can generally be trusted on important issues
    & Inst.\ Trust ($I$) & $+$ \\[3pt]
  Electoral fraud is a significant problem in modern democracies
    & Conspiracy ($C$) & $-$ \\
  \bottomrule
\end{longtable}

\sechead{Technology \& AI}
\begin{longtable}{p{0.68\textwidth} p{0.16\textwidth} c}
  \toprule
  \textbf{Statement} & \textbf{Dimension} & \textbf{Dir.}\\
  \midrule\endhead
  Artificial intelligence poses existential risks to humanity
    & Sci.\ Epist.\ ($S$) & $+$\,\textsuperscript{*} \\[3pt]
  Social media algorithms manipulate public opinion
    & Conspiracy ($C$) & $-$ \\[3pt]
  Technology companies should face stronger regulation
    & Inst.\ Trust ($I$) & $+$\,\textsuperscript{*} \\
  \bottomrule
\end{longtable}
\noindent\textsuperscript{*}Technology items included for attitudinal 
breadth; assigned reduced weight in $\varphi$ computation pending 
psychometric validation.

\sechead{Science \& Epistemology}
\begin{longtable}{p{0.68\textwidth} p{0.16\textwidth} c}
  \toprule
  \textbf{Statement} & \textbf{Dimension} & \textbf{Dir.}\\
  \midrule\endhead
  Scientific consensus is generally reliable
    & Sci.\ Epist.\ ($S$) & $+$ \\[3pt]
  Personal experience is as valid as scientific studies
    & Alt.\ Epist.\ ($E$) & $-$ \\[3pt]
  Experts often have hidden agendas
    & Conspiracy ($C$) & $-$ \\
  \bottomrule
\end{longtable}

\noindent\textbf{Scoring.}
Dimension scores ($I$, $S$, $C$, $E$) were computed as the mean of their 
respective items after normalisation to $[0,1]$.
The raw phase score was computed per Equation~\eqref{eq:phi-raw} 
and rescaled to $[0,\pi]$ per Equation~\eqref{eq:phi-scaled}.

\subsection*{Part B: Source Trust Assessment}
\label{app:partB}

Participants rated their trust in each information source on a five-point 
scale: \emph{No trust at all} --- \emph{Little trust} --- 
\emph{Moderate trust} --- \emph{High trust} --- \emph{Complete trust}.
These ratings were used as supplementary indicators of institutional trust 
and as qualitative validation for the $I$ dimension of $\varphi$.

\begin{longtable}{p{0.72\textwidth} p{0.22\textwidth}}
  \toprule
  \textbf{Source} & \textbf{Category}\\
  \midrule\endhead
  Academic research articles and scientific publications
    & Institutional ($I$) \\[3pt]
  Major Austrian and German-language newspapers 
  (\textit{Der Standard}, \textit{Die Presse}, \textit{Süddeutsche Zeitung})
    & Institutional ($I$) \\[3pt]
  Government health agencies (AGES, WHO)
    & Institutional ($I$) \\[3pt]
  Alternative health websites
    & Alternative ($E$) \\[3pt]
  Independent investigative journalists
    & Mixed \\[3pt]
  Social media influencers
    & Alternative ($E$) \\[3pt]
  University researchers and professors
    & Institutional ($I$) \\[3pt]
  Corporate press releases (from companies and businesses)
    & Conspiracy ($C$) \\[3pt]
  Peer-reviewed meta-analyses (comprehensive studies analysing 
  multiple research papers)
    & Institutional ($I$) \\[3pt]
  Personal testimony and anecdotes (stories from friends, family, 
  or online)
    & Alternative ($E$) \\
  \bottomrule
\end{longtable}

\subsection*{Part C: Open-Ended Questions}
\label{app:partC}

Participants were asked to respond to the following three questions in 
free text (minimum: a few sentences each).
Responses were not used in quantitative analyses but were examined 
qualitatively to validate cluster assignments and identify edge cases.

\begin{enumerate}
  \item Describe a time when you changed your mind about something 
        important. What caused the change?
  \item What makes you trust or distrust a source of information?
  \item When you encounter a claim you are unsure about, what do you do?
\end{enumerate}

\subsection*{Part D: Demographics and Individual Differences}
\label{app:partD}

\subsubsection*{D.1 Demographics}

\begin{itemize}
  \item Age (in years; open numeric field)
  \item Gender (Male / Female / Non-binary / Prefer to self-describe / 
        Prefer not to say)
  \item Political orientation (7-point scale: Very Left-wing --- 
        Very Right-wing)
\end{itemize}

\subsubsection*{D.2 Cognitive Reflection Test (CRT)}

The following five items assessed analytical versus intuitive reasoning.
Items 1--4 are the standard CRT \citep{frederick2005}; item~5 is 
a numeracy extension:

\begin{enumerate}
  \item Determine the value of $x$ that makes $x + 12 = 11$ true.
        \hfill \textit{(correct: $-1$)}
  \item A bat and a ball cost \texteuro1.10 in total. The bat costs 
        \texteuro1.00 more than the ball. How much does the ball cost?
        \hfill \textit{(correct: 5 cents)}
  \item If it takes 5 machines 5 minutes to make 5 widgets, how long 
        would it take 100 machines to make 100 widgets?
        \hfill \textit{(correct: 5 minutes)}
  \item In a lake, there is a patch of lily pads. Every day the patch 
        doubles in size. If it takes 48 days to cover the entire lake, 
        how long would it take to cover half the lake?
        \hfill \textit{(correct: 47 days)}
  \item Starting with one 10\,cm lily pad that doubles in area each day, 
        after how many days would it cover the entire Earth?
        \hfill \textit{(estimation; no single correct answer)}
\end{enumerate}

\subsubsection*{D.3 Need for Cognition (NFC)}

Six items from the short-form Need for Cognition scale 
\citep{cacioppo1984}, rated on the standard five-point agreement scale:

\begin{enumerate}
  \item I would prefer complex to simple problems.
  \item I like to have the responsibility of handling a situation that 
        requires a lot of thinking.
  \item Thinking is not my idea of fun. \hfill\textit{(reverse-scored)}
  \item I would rather do something that requires little thought than 
        something that is sure to challenge my thinking abilities.
        \hfill\textit{(reverse-scored)}
  \item I really enjoy a task that involves coming up with new solutions 
        to problems.
  \item I would prefer a task that is intellectual, difficult, and 
        important to one that is somewhat important but does not require 
        much thought.
\end{enumerate}

\subsubsection*{D.4 Actively Open-Minded Thinking (AOT)}

Six items assessing epistemic flexibility and openness to revision 
\citep{stanovich1997}, rated on the standard five-point agreement scale:

\begin{enumerate}
  \item People should take into consideration evidence that goes against 
        their beliefs.
  \item It is important to persevere in your beliefs even when evidence 
        is brought to bear against them. \hfill\textit{(reverse-scored)}
  \item Changing your mind is a sign of weakness. 
        \hfill\textit{(reverse-scored)}
  \item Intuition is the best guide in making decisions.
        \hfill\textit{(reverse-scored)}
  \item A person should always consider new possibilities.
  \item There are two kinds of people in this world: those who are for 
        the truth and those who are against it.
        \hfill\textit{(reverse-scored)}
\end{enumerate}

\section{Session~1 Experimental Instruments (Claim~1: 5G Health Effects)}
\label{app:session1}

The following three instruments were administered in Session~1, one per 
experimental group, all targeting the same claim:

\medskip
\noindent\textit{Claim~1 (false):} ``5G mobile communications causes 
significant health damage in humans.''\\
(\textit{German original:} ``5G-Mobilfunk verursacht signifikante 
gesundheitliche Schäden beim Menschen.'')

\medskip
\noindent All participants in all three conditions completed the same 
pre-intervention rating (truth judgement + confidence, 0--10 scale) and 
post-intervention rating on the same scales.
The intervention material differed by condition.
The same structure was repeated for Claims~2 and~3 in later sessions.

\subsection*{Session~1A --- Group~A: Traditional Fact-Checking}
\label{app:s1a}

\noindent\textbf{Timing:} $\approx$\,6 minutes total.

\medskip
\noindent\textbf{A1. Pre-intervention rating} ($\approx$1 min)
\begin{itemize}
  \item[A1.1] Truth judgement: ``How true is this statement in your 
    opinion?'' (0 = completely false \ldots 10 = completely true)
  \item[A1.2] Confidence: ``How certain are you?'' 
    (0 = very uncertain \ldots 10 = very certain)
\end{itemize}

\medskip
\noindent\textbf{A2. Evidence Pack} ($\approx$3 min, read only)

Participants read a structured summary of scientific consensus 
(English translation of German original):

\begin{quote}
\textit{Physical basis:} 5G uses non-ionising electromagnetic radiation 
that cannot damage DNA. Its energy lies well below biologically effective 
thresholds.

\textit{Health studies:} Large systematic reviews and longitudinal studies 
find no association between mobile network exposure and health damage.

\textit{Comparison with limits:} 5G exposure lies far below international 
safety limits.

\textit{Institutional assessment:} WHO, national health authorities, and 
professional associations see no evidence-based health risks.

\textit{Summary conclusion:} There is no scientific evidence that 5G causes 
health damage.
\end{quote}

\medskip
\noindent\textbf{A3. Comprehension check} ($\approx$1 min)
\begin{itemize}
  \item[A3.1] ``Which statement is correct?''\\
    5G radiation is ionising /
    5G radiation can directly damage DNA /
    \underline{5G radiation is non-ionising} /
    There are no physical differences to X-ray radiation
  \item[A3.2] ``What is correct according to the evidence?''\\
    Many studies show severe effects /
    Evidence is contradictory /
    \underline{Studies show no association} /
    There are no studies
\end{itemize}
(Correct answers underlined; incorrect selections triggered a prompt to 
re-read the evidence pack before proceeding.)

\medskip
\noindent\textbf{A4. Post-intervention rating} ($\approx$30 sec)
\begin{itemize}
  \item[A4.1] Truth judgement (after reading the scientific consensus 
    summary), same 0--10 scale
  \item[A4.2] Confidence (after reading), same 0--10 scale
\end{itemize}

\medskip
\noindent\textbf{A5. Brief justification} ($\approx$1 min)\\
``Why did you change your assessment --- or not?'' (1--2 sentences)

\subsection*{Session~1B --- Group~B: Phase-Aware Coherence Analysis}
\label{app:s1b}

\noindent\textbf{Timing:} $\approx$\,12 minutes total.

\medskip
\noindent\textbf{B1. Pre-intervention rating} ($\approx$1 min)\\
Identical structure to A1 (truth judgement + confidence, 0--10).

\medskip
\noindent\textbf{B2. Multi-perspective description} ($\approx$3 min, 
read only)

Participants read descriptions of three epistemic perspectives on the 
claim.
The descriptions were explicitly framed as non-evaluative summaries of 
typical argumentative approaches, not as endorsements:

\begin{description}
  \item[Perspective~1: Physics and Electrical Engineering.]
    This perspective examines electromagnetic radiation via physical 
    properties such as frequency, energy, and interaction with matter.
    5G uses non-ionising radio waves whose energy lies well below known 
    biological effect thresholds.
    A direct damage mechanism is regarded as physically implausible.

  \item[Perspective~2: Medical and Epidemiological Research.]
    This perspective evaluates health effects using controlled studies, 
    population data, and statistical associations.
    Large systematic reviews and longitudinal studies find no robust 
    association between mobile network exposure and health damage.
    Some subjective symptoms are reported but do not correlate 
    consistently with objective measurements.

  \item[Perspective~3: Critical or Alternative Health Narratives.]
    This perspective emphasises possible unknown risks of new technologies 
    and expresses scepticism toward official assessments.
    It frequently draws on individual cases, personal experience, or the 
    assumption that long-term effects have not been sufficiently studied.
    Institutional statements are sometimes perceived as interest-driven.
\end{description}

Participants then completed three open-response items (B3.1--B3.3):
for each perspective, they formulated one sentence summarising how that 
perspective would evaluate the claim in their own words.

\medskip
\noindent\textbf{B4. Structure analysis} ($\approx$2 min)
\begin{itemize}
  \item[B4.1] \textit{Shared ground:} ``Is there a point on which all 
    perspectives at least partially agree?'' (open text; e.g.\ 
    uncertainties, measurement problems, open questions)
  \item[B4.2] \textit{Central difference:} ``Where, in your opinion, lies 
    the decisive conflict between the perspectives?'' (open text; e.g.\ 
    mechanism, evidentiary standard, trust, time dimension)
\end{itemize}

\medskip
\noindent\textbf{B5. Coherence and isolation check} ($\approx$2 min)
\begin{itemize}
  \item[B5.1] \textit{Assumption audit:} ``What additional assumptions would 
    be necessary for the statement to be definitively true?'' (open text; 
    e.g.\ unknown physical effects, systematic mismeasurement, 
    long-term concealment)
  \item[B5.2] \textit{Structural classification:} ``Would you now say 
    this statement is \ldots'' (forced choice):\\
    (a) Well compatible with multiple perspectives\\
    (b) Contested, but in principle integrable\\
    (c) Strongly dependent on a single perspective\\
    (d) Structurally isolated (requires rejection of established 
        foundations)
\end{itemize}

\medskip
\noindent\textbf{B6. Post-intervention rating} ($\approx$30 sec)
\begin{itemize}
  \item[B6.1] Truth judgement (own re-evaluation), 0--10 scale
  \item[B6.2] Confidence (re-evaluation), 0--10 scale
  \item[B6.3] Brief justification: ``Why did you change your 
    assessment --- or not?'' (1--2 sentences)
\end{itemize}

\medskip
\noindent\textbf{B7. Brief reflection} (multiple choice, 
$\approx$30 sec)\\
``What was most decisive for your judgement?''
(Agreement or disagreement between perspectives /
Absence of a plausible mechanism /
Necessity of additional assumptions /
Own prior experience or intuition)

\subsection*{Session~1C --- Group~C: Control (Self-Reflection)}
\label{app:s1c}

\noindent\textbf{Timing:} $\approx$\,10 minutes total.\\
\noindent\textbf{Condition label:} \textit{Selbstreflexion ohne externes 
Material} (Self-reflection without external material).\\
Participants were explicitly instructed not to use any external sources.

\medskip
\noindent\textbf{C1. Pre-intervention rating} ($\approx$1 min)\\
Identical structure to A1 (truth judgement + confidence, 0--10).

\medskip
\noindent\textbf{C2. Thinking task} ($\approx$7 min)

Participants were asked to reconsider the claim and answer the 
following solely on the basis of their own knowledge and reasoning:
\begin{itemize}
  \item[C2.1] \textit{Arguments for:} ``What reasons speak, in your 
    opinion, \emph{for} the statement?'' (minimum 2--3 bullet points)
  \item[C2.2] \textit{Arguments against:} ``What reasons speak, in your 
    opinion, \emph{against} the statement?'' (minimum 2--3 bullet points)
  \item[C2.3] \textit{Uncertainties:} ``What makes you uncertain about 
    this statement?'' (minimum 1--2 sentences)
\end{itemize}

\medskip
\noindent\textbf{C3. Post-intervention rating} ($\approx$2 min)\\
Truth judgement and confidence re-assessment on the same 0--10 scales 
(items C3.1 and C3.2).

\medskip
\noindent\textbf{C4. Brief justification} ($\approx$30 sec)\\
``Why did you change your assessment --- or not?'' (1--2 sentences)

\bigskip
\noindent\textbf{Structural comparison of the three conditions.}
Table~\ref{tab:session-structure} summarises the key design differences.

\begin{table}[h!]
\centering
\caption{Structural comparison of Session~1 instruments across conditions.}
\label{tab:session-structure}
\begin{tabular}{llll}
  \toprule
  \textbf{Phase} & \textbf{Group~A (Fact-check)} 
    & \textbf{Group~B (Phase-aware)} 
    & \textbf{Group~C (Control)} \\
  \midrule
  Pre-rating     & Truth + confidence & Truth + confidence 
    & Truth + confidence \\
  Intervention   & Evidence pack      & 3 perspectives + 
    & Self-generated \\
                 & (consensus summary)& structure analysis 
    & pro/con arguments \\
  Fidelity check & Comprehension quiz & Perspective 
    & None \\
                 &                    & articulation (B3) & \\
  Post-rating    & Truth + confidence & Truth + confidence 
    & Truth + confidence \\
  Reflection     & Justification      & Justification + 
    & Justification \\
                 &                    & decision factors & \\
  Duration       & $\approx$6 min     & $\approx$12 min 
    & $\approx$10 min \\
  \bottomrule
\end{tabular}
\end{table}

\section{Session~2 Experimental Instruments (Claim~2: Urban Trees and Air Quality)}
\label{app:session2}

The following three instruments were administered in Session~2, one per 
experimental group, all targeting the same claim:

\medskip
\noindent\textit{Claim~2 (true but heavily qualified):} ``Trees in urban 
areas significantly reduce local air pollution.''\\
(\textit{German original:} ``B\"{a}ume in st\"{a}dtischen Gebieten reduzieren 
die lokale Luftverschmutzung signifikant.'')

\medskip
\noindent Unlike Claim~1 (clearly false), Claim~2 is \emph{partially 
true}: trees bind particulate matter and absorb some gaseous pollutants, 
but the effect is modest, highly context-dependent, and can be 
counterproductive in narrow street canyons.
This nuance was built into both the evidence pack (Session~2A) and the 
perspective descriptions (Session~2B), making Claim~2 a stronger test 
of whether the phase-aware approach preserves epistemic complexity 
rather than simply nudging participants toward rejection.

\subsection*{Session~2A --- Group~A: Traditional Fact-Checking}
\label{app:s2a}

\noindent\textbf{Timing:} $\approx$\,6 minutes total.

\medskip
\noindent\textbf{A1. Pre-intervention rating} ($\approx$30 sec)\\
Truth judgement (A1.1) and confidence (A1.2), 0--10 scale.

\medskip
\noindent\textbf{A2. Evidence Pack} ($\approx$3 min, read only)

\begin{quote}
\textit{Mechanisms:} Trees can bind particulate matter on leaves and 
absorb gaseous pollutants (e.g.\ NO$_2$) to a limited extent.

\textit{Measurements:} Studies show locally measurable effects, usually 
in the range of a few percentage points.
Traffic remains the dominant source of urban air pollution.

\textit{Context dependence:} In open street spaces, green infrastructure 
can help.
In narrow street canyons, however, it can reduce air circulation and 
worsen local pollution.

\textit{Institutional assessment:} Urban greening is recommended as a 
supplementary measure, not as the primary solution for air quality 
management.

\textit{Summary conclusion:} Trees can contribute to improving air 
quality, but the effect is limited and highly context-dependent.
\end{quote}

\noindent\textit{Note:} Unlike Session~1A, this evidence pack does not 
straightforwardly reject the claim.
It presents a qualified endorsement, acknowledging a real mechanism 
while stressing its limited magnitude and strong context-dependence.

\medskip
\noindent\textbf{A3. Comprehension check} ($\approx$1 min)
\begin{itemize}
  \item[A3.1] ``Which statement is most accurate?'' (single choice)\\
    Trees reduce air pollution strongly everywhere /
    \underline{Trees can help locally but are not the primary solution} /
    Trees have no measurable effect /
    There are no studies on this
  \item[A3.2] ``Which conditions most strongly influence the effect?''
    (multiple choice; all correct: traffic volume /
    street geometry / tree species and leaf area / weather and wind)
\end{itemize}
(Correct answer to A3.1 underlined.
A3.2 accepts multiple responses and does not trigger a re-read prompt.)

\medskip
\noindent\textbf{A4. Post-intervention rating} ($\approx$30 sec)\\
Truth judgement (A4.1) and confidence (A4.2), same 0--10 scales.

\medskip
\noindent\textbf{A5. Brief justification} ($\approx$1 min)\\
``Why did you change your assessment --- or not?'' (1--2 sentences)

\subsection*{Session~2B --- Group~B: Phase-Aware Coherence Analysis}
\label{app:s2b}

\noindent\textbf{Timing:} $\approx$\,9 minutes total.

\medskip
\noindent\textbf{B1. Pre-intervention rating} ($\approx$30 sec)\\
Truth judgement (B1.1) and confidence (B1.2), 0--10 scale.

\medskip
\noindent\textbf{B2. Multi-perspective description} ($\approx$2 min, read only)

Three perspectives were presented as compact bullet-point summaries 
(more concise than Session~1B's prose paragraphs):

\begin{description}
  \item[Perspective~1: Air Quality Research.]
    Measures pollutants directly (fine particulates, NO$_2$);
    mostly finds small, local effects;
    identifies traffic as the primary problem.

  \item[Perspective~2: Urban Climatology.]
    Investigates air flows and street geometry;
    effect strongly dependent on building form and wind;
    in narrow streets, sometimes counterproductive effects.

  \item[Perspective~3: Public Perception / Urban Policy.]
    Associates green spaces with higher quality of life;
    frequently conflates heat, noise, and air quality;
    uses simplified messaging (``more trees = clean air'').
\end{description}

Participants articulated each perspective in one sentence (items B3.1--B3.3).

\medskip
\noindent\textbf{B4. Structure analysis} ($\approx$2 min)
\begin{itemize}
  \item[B4.1] \textit{Shared ground} (open text; e.g.\ uncertainties,
    measurement problems, the word `significantly')
  \item[B4.2] \textit{Central difference} (open text; e.g.\ measurement
    variable, context, expectations)
\end{itemize}

\medskip
\noindent\textbf{B5. Bridge question} ($\approx$1 min)\\
``What concrete information or measurement could help to better 
reconcile the different perspectives?''\\
(open text; e.g.\ specific street types, comparison measurements, time periods)

\noindent\textit{Note:} This item replaced the isolation classification 
(B5.2 in Session~1B). For a partially true claim, a constructive 
bridge-building prompt is more appropriate than a forced structural 
categorisation.

\medskip
\noindent\textbf{B6. Post-intervention rating} ($\approx$30 sec)\\
Truth judgement (B6.1) and confidence (B6.2), 0--10 scales.

\medskip
\noindent\textbf{Brief justification} (1--2 sentences).

\noindent\textit{Note:} The decision-factor checklist (B7 in Session~1B) 
was omitted in Session~2B.

\subsection*{Session~2C --- Group~C: Control (Self-Reflection)}
\label{app:s2c}

\noindent\textbf{Timing:} $\approx$\,10 minutes total. Identical in 
structure to Session~1C; only the claim text changed.

\medskip
\noindent\textbf{C1.} Pre-intervention rating (truth + confidence, 0--10).\\
\noindent\textbf{C2.} Thinking task ($\approx$7 min): arguments for (C2.1),
arguments against (C2.2), uncertainties (C2.3); 
no external sources or aids permitted.\\
\noindent\textbf{C3.} Post-intervention rating (truth + confidence).\\
\noindent\textbf{C4.} Brief justification (1--2 sentences).

\bigskip
\noindent\textbf{Design differences between Sessions~1 and~2.}
Table~\ref{tab:s2diff} summarises the key methodological changes 
introduced for the nuanced Claim~2.

\begin{table}[h!]
\centering
\small
\caption{Key design differences between Session~1 (Claim~1: false)
and Session~2 (Claim~2: true but qualified).}
\label{tab:s2diff}
\begin{tabular}{p{0.21\textwidth}p{0.35\textwidth}p{0.35\textwidth}}
  \toprule
  \textbf{Feature} & \textbf{Session~1} & \textbf{Session~2} \\
  \midrule
  Claim truth value & Clearly false 
    & Partially true; heavily qualified \\[3pt]
  Evidence pack tone (Group~A) 
    & Consensus rejection 
    & Qualified endorsement with caveats \\[3pt]
  Comprehension check (A3) 
    & Binary right/wrong; re-read on failure 
    & Nuanced best-fit + multi-select moderators;
      no re-read prompt \\[3pt]
  Perspective format (Group~B)
    & Prose paragraphs ($\approx$3 min) 
    & Compact bullet points ($\approx$2 min) \\[3pt]
  Post-perspective task (Group~B, B5) 
    & Isolation classification (4-option forced choice) 
    & Bridge question (open text) \\[3pt]
  Decision-factor checklist (Group~B, B7) 
    & Included & Omitted \\
  \bottomrule
\end{tabular}
\end{table}

\section{Session~3 Experimental Instruments (Claim~3: Vaccines and the Human Genome)}
\label{app:session3}

The following three instruments were administered in Session~3, one per 
experimental group.
Session~3 serves as the transfer and retention test of the study: 
participants applied whatever strategy or disposition they had developed 
across Sessions~1 and~2 to a new, high-stakes claim they had not 
encountered before.

\medskip
\noindent\textit{Claim~3 (false):} ``Vaccines alter the human genetic 
material.''\\
(\textit{German original:} ``Impfstoffe ver\"{a}ndern das menschliche Erbgut.'')

\medskip
\noindent Claim~3 is clearly false but is among the most consequential 
and emotionally loaded misinformation claims in contemporary public health 
discourse, having circulated widely during the COVID-19 vaccination 
campaign.
Like Claim~1 (5G), it admits no partial truth: neither mRNA vaccines 
nor classical vector vaccines enter the cell nucleus, and no approved 
vaccine integrates into the genome.
The claim's persistence despite clear refutation makes it a strong test 
of whether phase-aware coherence training generalises beyond the original 
training domain.

\medskip
\noindent\textit{Note on labelling.}
Session~3B was internally labelled \textit{Perspektivwechsel} 
(Perspective Shift) on its cover page, in contrast to Sessions~1B and~2B 
which carried no subtitle.
This labelling difference was not intended as an experimental manipulation 
and is noted here for completeness.

\subsection*{Session~3A --- Group~A: Traditional Fact-Checking}
\label{app:s3a}

\noindent\textbf{Timing:} $\approx$\,6 minutes total.\\
\noindent\textit{Cover page label:} \textit{Session 3A Evidence Fakten.}

\medskip
\noindent\textbf{A1. Pre-intervention rating}\\
Truth judgement (A1.1) and confidence (A1.2), 0--10 scale.

\medskip
\noindent\textbf{A2. Evidence Pack} ($\approx$3 min, read only)

\begin{quote}
\textit{Biological basis:} Human genetic material (DNA) is located in 
the cell nucleus.
Vaccines---including mRNA vaccines---do not enter the cell nucleus and 
therefore cannot alter DNA.

\textit{mRNA vaccines:} mRNA serves only as a template for producing a 
protein and is degraded after a short time in the cytoplasm of the cell.
It is not reverse-transcribed into DNA and is not integrated into the 
genome.

\textit{Classical vaccines:} Inactivated or vector vaccines contain 
either harmless components of a pathogen or attenuated viruses, neither 
of which causes any alteration of the human genome.

\textit{Scientific consensus:} International health organisations and 
professional associations agree that approved vaccines do not alter 
the human genome.

\textit{Summary conclusion:} The claim that vaccines alter the genome 
is scientifically false.
\end{quote}

\noindent\textit{Note:} The evidence pack for Claim~3 is the most 
mechanistically explicit of the three sessions: it specifies the 
subcellular localisation of DNA, addresses both mRNA and classical 
vaccine types separately, and names the reverse-transcription step 
that would be required (and does not occur).
This level of detail reflects the technical character of the claim, 
which hinges on cellular biology rather than epidemiology or physics.

\medskip
\noindent\textbf{A3. Comprehension check} ($\approx$1 min)
\begin{itemize}
  \item[A3.1] ``Which statement is correct?'' (single choice)\\
    Vaccines alter the DNA in the cell nucleus /
    mRNA vaccines enter the cell nucleus /
    \underline{Vaccines work without altering the genome} /
    There are no established findings on this
  \item[A3.2] ``What happens to the mRNA after vaccination?''
    (single choice)\\
    It is permanently converted into DNA /
    It remains in the body for life /
    \underline{It is degraded after a short time} /
    It alters the genome indirectly
\end{itemize}
(Correct answers underlined.)

\medskip
\noindent\textbf{A4. Post-intervention rating}\\
Truth judgement (A4.1) and confidence (A4.2), same 0--10 scales,
labelled ``after reading the scientific consensus summary''.

\medskip
\noindent\textbf{A5. Brief justification} ($\approx$1 min)\\
``Why did you change your assessment --- or not?'' (1--2 sentences)

\subsection*{Session~3B --- Group~B: Phase-Aware Coherence Analysis}
\label{app:s3b}

\noindent\textbf{Timing:} $\approx$\,9 minutes total.

\medskip
\noindent\textbf{B1. Pre-intervention rating}\\
Truth judgement (B1.1) and confidence (B1.2), 0--10 scale.

\medskip
\noindent\textbf{B2. Multi-perspective description} ($\approx$3 min, 
read only)

Three perspectives were presented in a prose-paragraph format 
(returning to the more detailed style of Session~1B rather than 
the compact bullets of Session~2B):

\begin{description}
  \item[Perspective~1: Molecular Biology / Genetics.]
    This perspective knows the mechanisms of DNA replication, 
    understands that genomic integration requires highly specific 
    enzymes, and sees no plausible pathway for vaccine-derived 
    nucleic acids to enter the genome.

  \item[Perspective~2: Medicine / Immunology.]
    This perspective treats vaccines as training signals for the 
    immune system, focuses on efficacy and side effects, and 
    attributes genetic changes to other causes entirely.

  \item[Perspective~3: Public Debate / Disinformation.]
    This perspective uses terms such as ``gene manipulation'' 
    imprecisely, conflates genetic engineering, mRNA, and the 
    genome, and amplifies uncertainty through simplified or 
    false analogies.
\end{description}

\noindent\textit{Note:} Perspective~3 in Session~3B is explicitly 
characterised as a disinformation narrative---a stronger framing than 
in Sessions~1B and~2B, where the third perspective was presented as a 
``critical or alternative'' viewpoint without pejorative labelling.
This design choice was deliberate: for a claim with no epistemic 
legitimacy, presenting the dissenting position as a coherent 
``alternative health narrative'' (as in Session~1B) risks false balance.
Reviewers should note this asymmetry when comparing inter-session 
perspective articulation data.

Participants articulated each perspective in one sentence 
(items B3.1 and B3.2; note that the third item was also labelled B3.2 
on the original REDCap form---a typographical error in the instrument).

\medskip
\noindent\textbf{B4 (labelled B3 on form). Structure analysis} 
($\approx$2 min)
\begin{itemize}
  \item[B4.1] \textit{Shared ground:} all perspectives partly agree 
    on\ldots{} (open text)
  \item[B4.2] \textit{Central difference:} decisive conflict between 
    perspectives (open text; e.g.\ biological mechanism, conceptual 
    understanding, institutional trust)
\end{itemize}

\medskip
\noindent\textbf{B5. Bridge question} ($\approx$1 min)\\
``What concrete information or representation could help to better 
reconcile the perspectives?''\\
(open text; e.g.\ schematic cell diagrams, comparison with real 
genetic modifications, long-term data)

\medskip
\noindent\textbf{B6. Post-intervention rating} ($\approx$30 sec)\\
Truth judgement (B6.1) and confidence (B6.2), 0--10 scales.

\medskip
\noindent\textbf{Brief justification} (B4 on form, 1--2 sentences).

\subsection*{Session~3C --- Group~C: Control (Self-Reflection)}
\label{app:s3c}

\noindent\textbf{Timing:} $\approx$\,10 minutes total.\\
\noindent\textit{Cover page label:} \textit{Session 3C Selbstreflexion 
OHNE Zusatzinformationen/Hilfestellungen} (Self-reflection WITHOUT 
additional information or guidance).\\
The ``WITHOUT'' was printed in all-caps in the original, emphasising 
the absence of external aids more strongly than in Sessions~1C and~2C.

\medskip
\noindent\textit{Structural note:} In Session~3C, the pre-intervention 
rating (C1.1 and C1.2) and the thinking task header (C2) appeared on 
the same survey page, unlike Sessions~1C and~2C where they were on 
separate pages.
The items themselves were unchanged.

\medskip
\noindent\textbf{C1. Pre-intervention rating}\\
Truth judgement (C1.1) and confidence (C1.2), 0--10 scale.

\medskip
\noindent\textbf{C2. Thinking task} ($\approx$7 min)\\
Same structure as Sessions~1C and~2C:
arguments for (C2.1), arguments against (C2.2, minimum 2--3 bullet 
points each), uncertainties (C2.3, minimum 1--2 sentences);
no external sources or aids.

\medskip
\noindent\textbf{C3. Post-intervention rating}\\
Truth judgement (C3.1) and confidence (C3.2), 0--10 scales.
(Labelled ``Teil C2 --- Zweite Einsch\"{a}tzung'' on the REDCap form, 
another minor numbering inconsistency in the instrument.)

\medskip
\noindent\textbf{C4. Brief justification} (1--2 sentences).

\bigskip
\noindent\textbf{Design continuity and deviations across all three 
sessions.}
Table~\ref{tab:allsessions} provides a consolidated overview of the 
complete three-session instrument sequence.

\begin{table}[h!]
\centering
\small
\caption{Consolidated overview of the three-session instrument design 
across all conditions and claims.}
\label{tab:allsessions}
\begin{tabular}{p{0.12\textwidth}p{0.17\textwidth}p{0.22\textwidth}
                 p{0.22\textwidth}p{0.17\textwidth}}
  \toprule
  & \textbf{Claim} & \textbf{Group A} 
    & \textbf{Group B} & \textbf{Group C} \\
  & & \textbf{(Fact-check)} & \textbf{(Phase-aware)} 
    & \textbf{(Control)} \\
  \midrule
  Session~1 
    & 5G health risk (false) 
    & Evidence pack; binary comprehension check; re-read prompt 
    & Prose perspectives; isolation classification (B5.2); 
      decision checklist (B7) 
    & Self-reflection; pro/con + uncertainties \\[4pt]
  Session~2 
    & Urban trees / air quality (partially true) 
    & Evidence pack; nuanced comprehension check; no re-read prompt 
    & Bullet perspectives; bridge question (B5); no B7 checklist 
    & Self-reflection; identical structure \\[4pt]
  Session~3 
    & Vaccines / genome (false) 
    & Evidence pack; binary comprehension check; re-read prompt 
    & Prose perspectives; disinformation label on P3; bridge question; 
      no B7 checklist 
    & Self-reflection; C1+C2 merged on one page \\
  \bottomrule
\end{tabular}
\end{table}
\section{Pre-Study Survey: Descriptive Analysis and Visualisations}
\label{app:prestudy-analysis}

All statistics and figures in this appendix are derived from the 
actual pre-study survey responses ($N=45$); no simulated or synthetic 
data are used.
The analyses establish (i)~the distributional properties of the 
epistemic phase coordinate $\varphi$, (ii)~dimension profiles that 
characterise each cluster, (iii)~source trust profiles, 
(iv)~individual-difference measures, and (v)~demographics.

\paragraph{Cluster label convention.}
The formula $\varphi_i^{(\text{raw})} = [(I_i+S_i)-(C_i+E_i)]/4$ 
produces positive scores for participants with high institutional 
trust ($I$) and scientific alignment ($S$), and negative scores 
for those with high conspiracy openness ($C$) and experiential 
epistemology ($E$).
After rescaling to $[0,\pi]$, \textbf{C1} ($\varphi \approx 0$, 
$n=5$) is the \emph{anti-institutional} cluster (low $I$, high $C$); 
\textbf{C3} ($\varphi \approx \pi$, $n=19$) is the 
\emph{pro-institutional} cluster (high $I$, low $C$); 
\textbf{C2} ($\varphi \approx \pi/2$, $n=21$) is the moderate cluster.
This is the opposite of informal expectations based on
``Cluster 1 = first = default'' intuition, and is the source of the 
label correction documented in the companion analysis file.

\subsection*{S1.~Phase Coordinate Distribution}

\begin{figure}[h!]
  \centering
  \includegraphics[width=\textwidth]{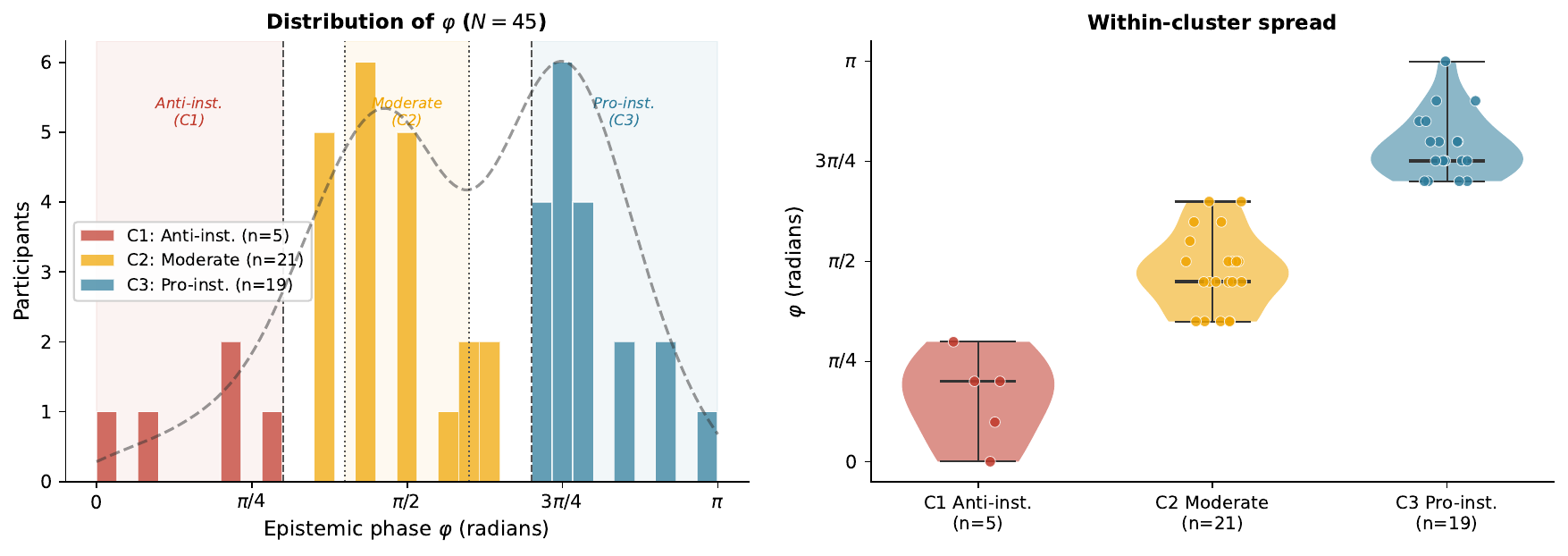}
  \caption{%
    \textbf{Left:} Histogram of $\varphi$ ($N=45$), colour-coded by 
    cluster. Dashed lines mark buffer-zone boundaries; the dashed 
    curve is a kernel-density estimate. C1 (red, $n=5$) clusters near 0 
    (anti-institutional); C3 (blue, $n=19$) clusters near $\pi$ 
    (pro-institutional). \textbf{Right:} Violin plots with jittered 
    individual data points.
  }
  \label{fig:phi-dist}
\end{figure}

The cluster sizes are $n_{\text{C1}}=5$, $n_{\text{C2}}=21$, 
$n_{\text{C3}}=19$. The small C1 cell ($n=5$) is the most important 
caveat for interpreting all C1-level results in the main paper.

Table~\ref{tab:phi-descriptives} reports summary statistics.

\begin{table}[h!]
\centering\small
\caption{Phase coordinate $\varphi$ and dimension means (SD) by cluster. 
         Dimensions normalised [0--1]; $\varphi$ in radians.}
\label{tab:phi-descriptives}
\begin{tabular}{lrrrrrrr}
  \toprule
  & & \multicolumn{4}{c}{\textbf{Dimension mean (SD)}}
    & \multicolumn{2}{c}{\textbf{$\varphi$}} \\
  \cmidrule(lr){3-6}\cmidrule(lr){7-8}
  \textbf{Cluster} & $n$ & $I$ & $S$ & $C$ & $E$ & $M$ & $SD$ \\
  \midrule
  C1 Anti-inst.    &  5 & 0.25 (0.23) & 0.47 (0.32) & 0.78 (0.19) & 0.80 (0.21) & 0.50 & 0.36 \\
  C2 Moderate      & 21 & 0.54 (0.16) & 0.75 (0.18) & 0.63 (0.16) & 0.71 (0.27) & 1.49 & 0.30 \\
  C3 Pro-inst.     & 19 & 0.65 (0.13) & 0.84 (0.18) & 0.41 (0.15) & 0.36 (0.24) & 2.48 & 0.25 \\
  \midrule
  Total            & 45 & & & & & 1.80 & 0.72 \\
  \bottomrule
\end{tabular}
\end{table}

$\varphi$ loadings: $r(I,\varphi)=+0.68$, $r(S,\varphi)=+0.45$, 
$r(C,\varphi)=-0.73$, $r(E,\varphi)=-0.62$ (all $p<.01$).

\clearpage
\subsection*{S2.~Dimension Profiles and Belief Items}

\begin{figure}[h!]
  \centering
  \includegraphics[width=\textwidth]{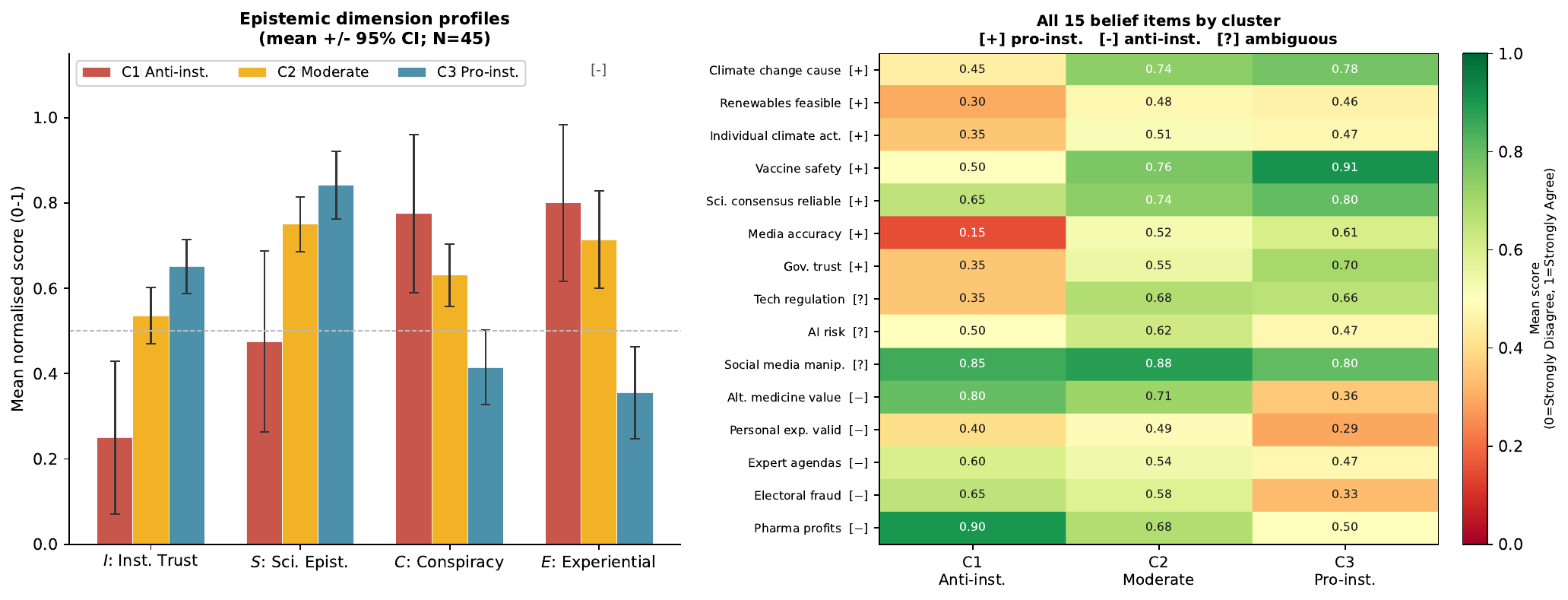}
  \caption{%
    \textbf{Left:} Dimension profiles ($I$, $S$, $C$, $E$) by cluster 
    (mean $\pm$ 95\,\% CI). $I$ and $S$ increase C1$\to$C3; $C$ and 
    $E$ decrease. \textbf{Right:} Heatmap of all 15 belief items by 
    cluster (green=agree, red=disagree; brackets = $\varphi$ 
    contribution direction).
  }
  \label{fig:dims}
\end{figure}

\clearpage
\subsection*{S3.~Intercorrelations}

\begin{figure}[h!]
  \centering
  \includegraphics[width=0.62\textwidth]{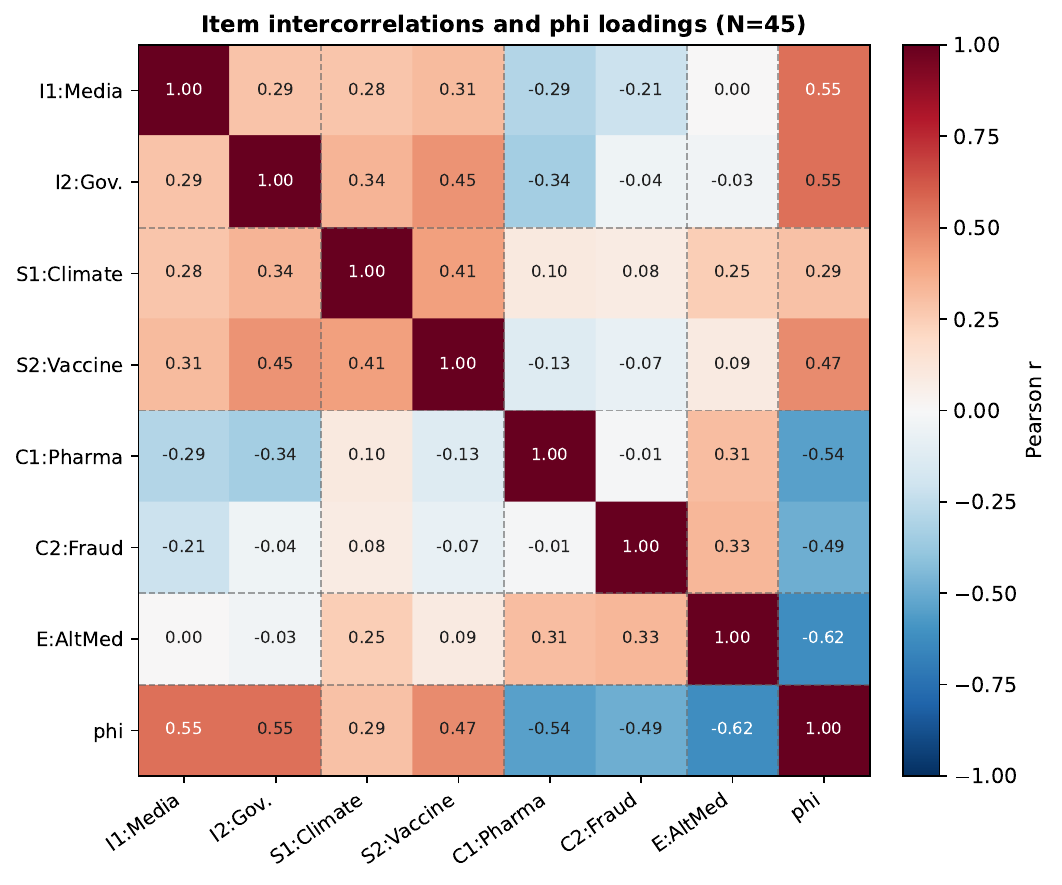}
  \caption{Pearson correlation matrix of phase-defining items and 
    $\varphi$ ($N=45$). Within-block correlations are positive; 
    cross-block $\{I,S\}$ vs.\ $\{C,E\}$ correlations are negative, 
    confirming the two construct poles are genuinely opposed.}
  \label{fig:corr}
\end{figure}

\clearpage
\subsection*{S4.~Source Trust Profiles}

\begin{figure}[h!]
  \centering
  \includegraphics[width=\textwidth]{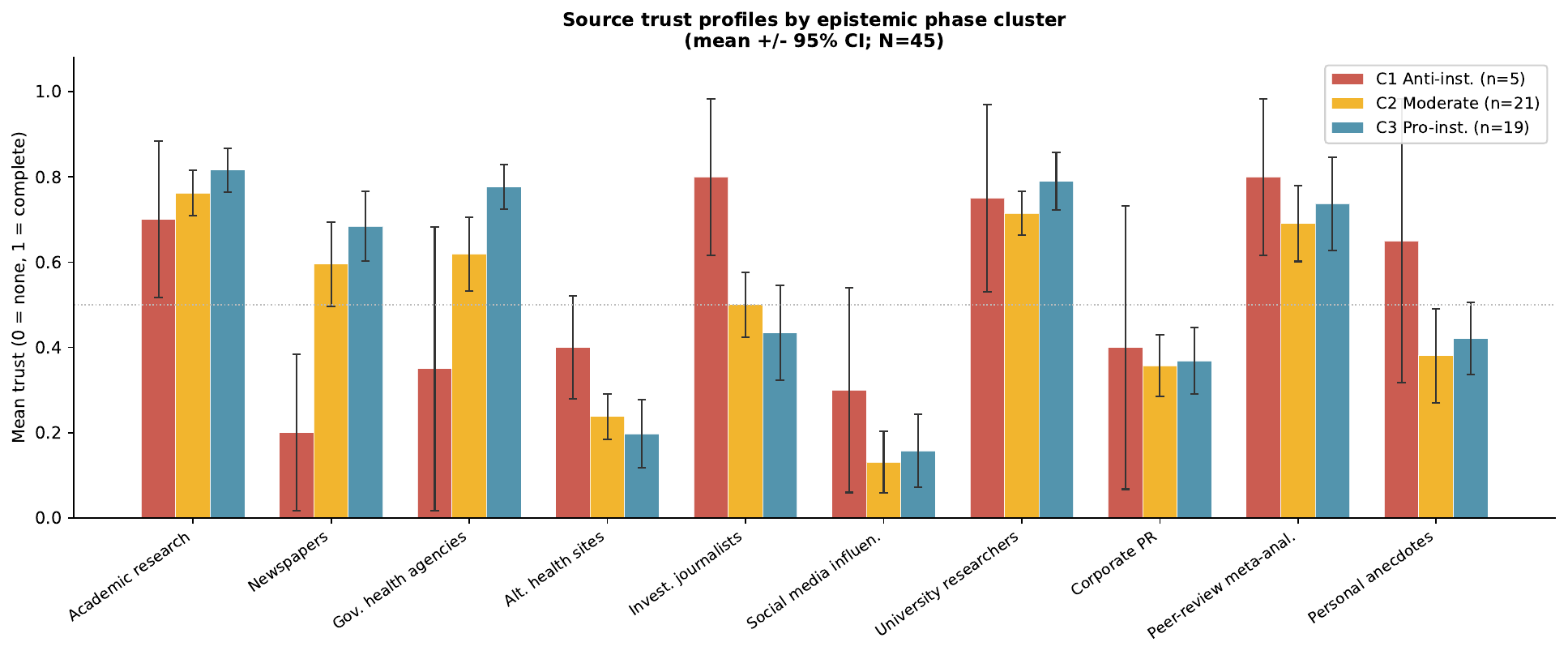}
  \caption{Mean source trust (0=no trust, 1=complete trust; 
    mean $\pm$ 95\,\% CI) by cluster. Institutional and scientific 
    sources increase C1$\to$C3; alternative health and anecdotal 
    sources show the opposite gradient, cross-validating the 
    dimension profiles from Part~A of the survey.}
  \label{fig:trust}
\end{figure}

\clearpage
\subsection*{S5.~Individual-Difference Scales}

\begin{figure}[h!]
  \centering
  \includegraphics[width=\textwidth]{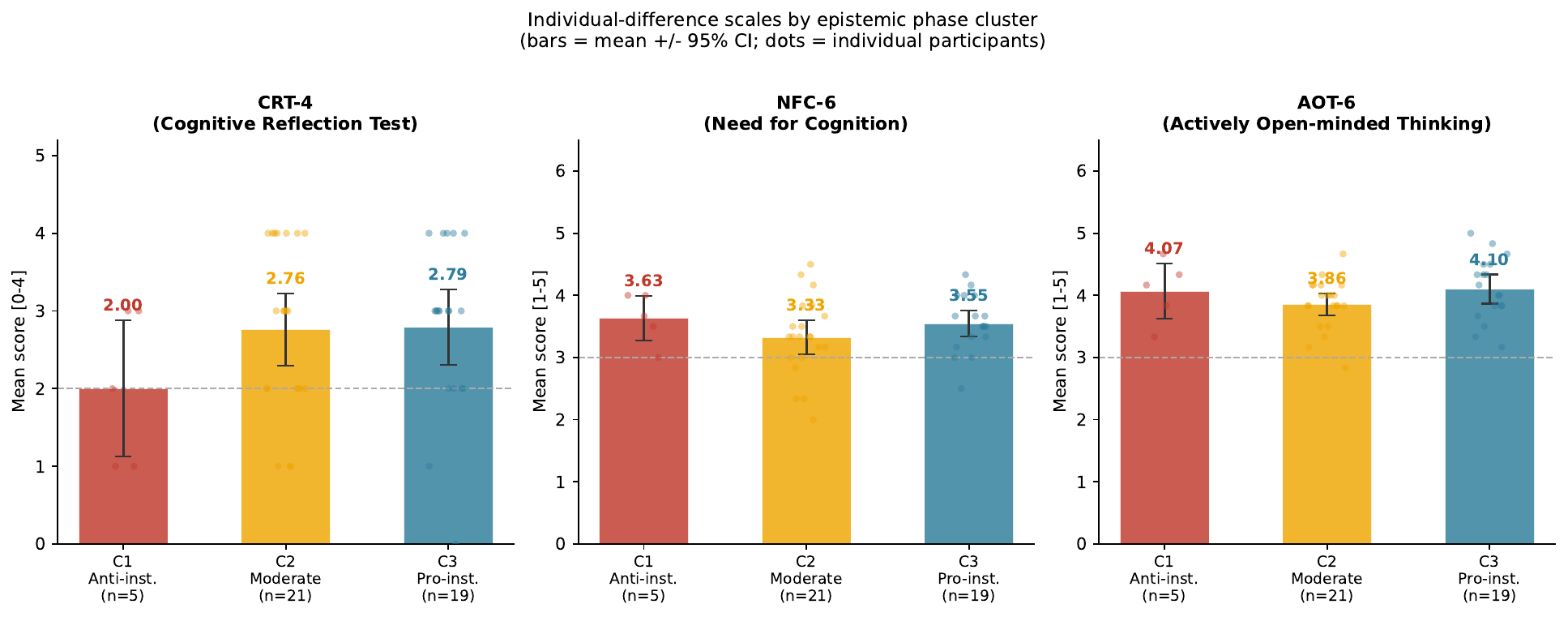}
  \caption{CRT-4, NFC-6, and AOT-6 by cluster (mean $\pm$ 95\,\% CI; 
    dashed line = scale midpoint; $n=44$ for NFC and AOT due to one 
    missing response). CRT and AOT do \emph{not} decrease 
    monotonically from C3 to C1, indicating that the 
    anti-institutional orientation in this sample is not primarily 
    driven by lower analytic capacity.}
  \label{fig:scales}
\end{figure}

\begin{table}[h!]
\centering\small
\caption{Individual-difference scale means (SD) by cluster 
         ($N=44$--$45$).}
\label{tab:scales}
\begin{tabular}{lrrrr}
  \toprule
  \textbf{Cluster} & CRT-4 [0--4] & NFC [1--5] & AOT [1--5] & $n$ \\
  \midrule
  C1 Anti-inst.  & 2.00 (1.00) & 3.63 (0.41) & 4.07 (0.51) &  5 \\
  C2 Moderate    & 2.76 (1.09) & 3.33 (0.63) & 3.86 (0.41) & 21 \\
  C3 Pro-inst.   & 2.79 (1.08) & 3.55 (0.46) & 4.10 (0.51) & 19 \\
  \midrule
  Total          & 2.69 (1.08) & 3.45 (0.55) & 3.98 (0.47) & 44 \\
  \bottomrule
\end{tabular}
\end{table}

CRT and AOT scores are near-identical for C2 and C3 
(CRT: 2.76 vs.\ 2.79; AOT: 3.86 vs.\ 4.10), and C1 shows 
the lowest CRT despite the highest AOT---though the $n=5$ C1 cell 
precludes reliable inference.
The near-zero correlations $r(\varphi, \text{CRT})=+0.20$ ($p=.20$) 
and $r(\varphi, \text{NFC})=-0.02$ ($p=.89$) indicate that 
$\varphi$ captures something largely orthogonal to general 
analytical capacity \citep{pennycook2019,douglas2019}.

\clearpage
\subsection*{S6.~Demographics}

\begin{figure}[h!]
  \centering
  \includegraphics[width=\textwidth]{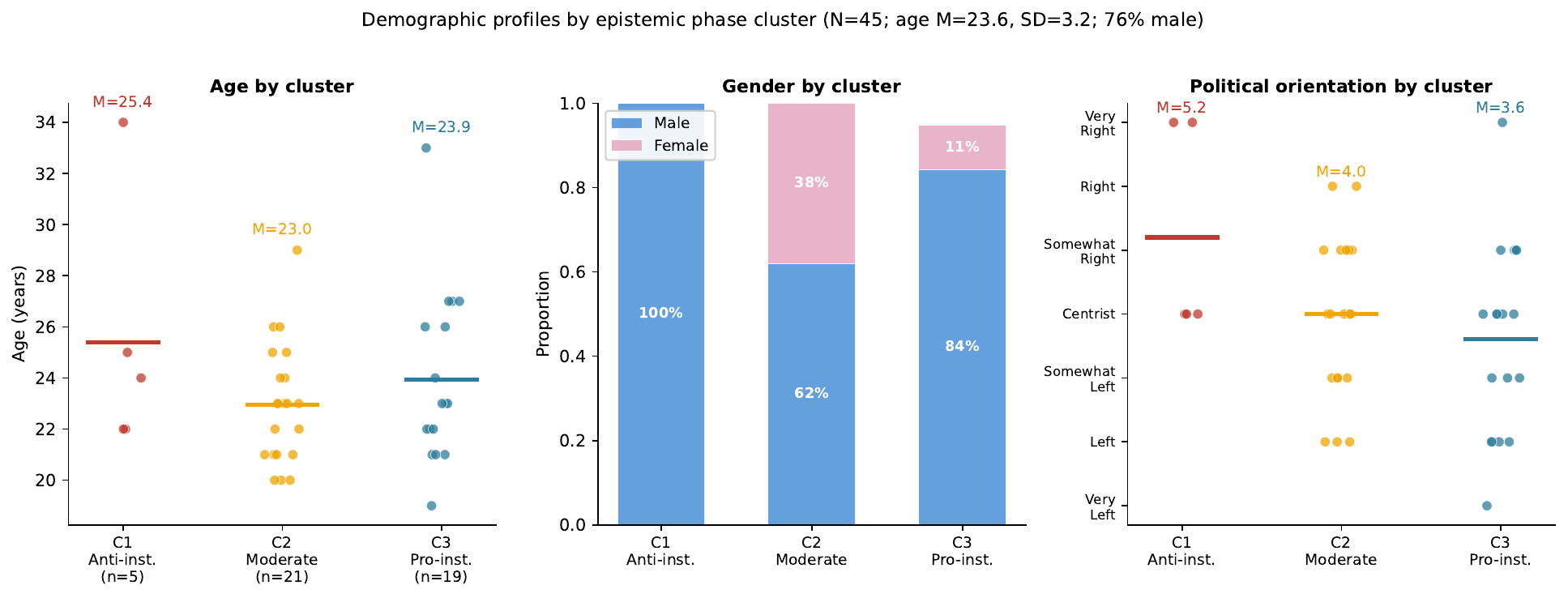}
  \caption{Demographic distributions by cluster ($N=45$; 
    one gender response missing). \textbf{Left:} Age (bars = means). 
    \textbf{Centre:} Gender proportions. 
    \textbf{Right:} Political orientation (7-point scale; bars = means). 
    The sample is predominantly male (34/45, 76\,\%) and young 
    ($M=23.6$, $SD=3.2$ yr). Cluster differences are small on all 
    three demographic variables.}
  \label{fig:demographics}
\end{figure}

\end{document}